\pdfoutput=1

\documentclass{JHEP3}

\usepackage{amsmath}
\usepackage{amssymb}
\usepackage{graphicx}
\usepackage{slashed}
\usepackage{youngtab}

\newcommand{\im}{{\rm Im\,}}

\newcommand{\bR}{\mathbb{R}}
\newcommand{\bZ}{\mathbb{Z}}

\newcommand{\cD}{\mathcal{D}}
\newcommand{\cE}{\mathcal{E}}
\newcommand{\cF}{\mathcal{F}}

\newcommand{\cM}{\mathcal{M}}
\newcommand{\cN}{\mathcal{N}}
\newcommand{\cR}{\mathcal{R}}
\newcommand{\cO}{\mathcal{O}}

\newcommand{\cV}{\mathcal{V}}

\newcommand{\ov}{\overline}
\newcommand{\del}{\partial}

\newcommand{\drawsquare}[2]{\hbox{%
\rule{#2pt}{#1pt}\hskip-#2pt
\rule{#1pt}{#2pt}\hskip-#1pt
\rule[#1pt]{#1pt}{#2pt}}\rule[#1pt]{#2pt}{#2pt}\hskip-#2pt
\rule{#2pt}{#1pt}}

\newcommand{\fund}{\raisebox{-.5pt}{\drawsquare{6.5}{0.4}}}
\newcommand{\Ysymm}{\raisebox{-.5pt}{\drawsquare{6.5}{0.4}}\hskip-0.4pt%
        \raisebox{-.5pt}{\drawsquare{6.5}{0.4}}}
\newcommand{\Yasymm}{\raisebox{-3.5pt}{\drawsquare{6.5}{0.4}}\hskip-6.9pt%
        \raisebox{3pt}{\drawsquare{6.5}{0.4}}}
\newcommand{\antifund}{\overline{\fund}}

\title{TASI Lectures: Particle Physics from Perturbative and Non-perturbative Effects
in D-braneworlds$^*$}

\author{Mirjam Cveti\v{c}$^{1,2}$ and James Halverson$^1$\\
  $^1$ Department of Physics and Astronomy, University of Pennsylvania,\\
  Philadelphia, PA 19104-6396, USA\\
  
$^2$ Center for Applied Mathematics and Theoretical Physics, \\
University of Maribor, Maribor, Slovenia

\vspace{.5cm}
  E-mail: \email{cvetic@cvetic.hep.upenn.edu}, \email{jhal@physics.upenn.edu}}

\abstract{In these notes 
we review aspects of semi-realistic particle physics from the point of
view of type II orientifold compactifications. We discuss the
appearance of gauge theories on spacetime filling D-branes which wrap non-trivial cycles in the Calabi-Yau. Chiral matter can appear at their intersections, with a natural interpretation of family replication given by the topological intersection number. We discuss global
consistency, including tadpole cancellation and the generalized Green-Schwarz
mechanism, and also the importance of related global $U(1)$ symmetries for
superpotential couplings. We review the basics of D-instantons, which can
generate superpotential corrections to charged
matter couplings forbidden by the global $U(1)$ symmetries and may play an important role in moduli stabilization.
Finally, for the purpose of studying the landscape, we discuss certain advantages of studying quiver gauge theories
which arise from type II orientifold compactifications rather than globally
defined models. We utilize the type IIa geometric picture and CFT techniques to illustrate the main physical points, though sometimes we supplement the discussion from the type IIb perspective using complex algebraic geometry.}

\preprint{UPR-1225-T}
\dedicated{$^*$ based on lectures given by M.C. at TASI 2010.}
\begin{document}

\section{Introduction}
Progress in our understanding of quantum field theory and particle physics over the last
fifty years has given us a truly remarkable model in the Standard Model of particle physics.
To high accuracy and precision, based on all experimental evidence to date, it appears to be
the correct low energy effective field theory for all particle interactions below the
weak scale. By now there is much excitement about the results which will come from the
Large Hadron Collider (LHC) over the next few years. Most expect that the Higgs boson will
be found, at which point all particles of the Standard Model will have been discovered.

Compared to other possibilities offered by non-abelian gauge theory, the Standard Model is rather complicated:
there are three families of quarks and leptons which transform under the gauge group $SU(3)_C\times SU(2)_L\times U(1)_Y$
and 26 parameters which make up the particle masses, mixing angles, and gauge coupling constants. There are large hierarchies
in the parameters, as masses of the lightest and heaviest fermions in the theory differ by over ten orders of magnitude.
Furthermore, though the up-flavor quarks (for example) transform in the same way with respect to the symmetries of
the quantum theory, there is a hierarchy of about five orders of magnitude between the masses of the $u$-quark and $t$-quark.
These rather striking experimental facts necessitate a theoretical explanation, and the Standard Model, though very successful,
does not provide one.

Any underlying theoretical framework should be able to explain the origin of the Standard Model gauge group,
particle representations, and parameters. If it is to be the fundamental theory of nature, that framework should also
give a sensible theory of quantum gravity. The best candidate for such a framework is superstring theory, which has been
shown to naturally give rise to all of these ingredients. 

Identifying particular string vacua
which could give rise to our world is often difficult, though. One difficulty is that superstring theory requires 
ten-dimensional spacetime, and therefore six of those dimensions must be compact
and of very small size to evade experimental bounds on extra dimensions. For the sake of $\cN=1$ supersymmetry, the standard type of manifold for compactification
is a Calabi-Yau threefold, of which there are at least thirty thousand\footnote{This is being as conservative as possible. In \cite{Kreuzer:2000xy}, 473,800,776 four (complex) dimensional toric varieties were identified which have a Calabi-Yau threefold hypersurface by Batyrev's construction. These
threefolds have 30,108 distinct pairs of Hodge numbers, giving the lower bound on the number of topologically distinct threefolds.}, and almost certainly many more. In addition to other
ingredients one can choose in defining a string theory, this choice of compactification manifold gives rise to a vast number
of string theories whose four-dimensional effective theory might possibly give rise to the particle physics seen in our world. These vacua are conjectured to be local minima of a potential on the moduli space of string theory, a notion which is often referred to as the landscape \cite{Susskind:2003kw,Denef:2004ze,Schellekens:2008kg}.

In these lectures, we focus on a corner of the string landscape which addresses many of the fundamental
questions in particle physics in an illuminating geometric fashion. We focus on type II string theory
(IIa in concrete examples), where spacetime-filling stacks of D-branes wrapping non-trivial cycles in the Calabi-Yau
give rise to four-dimensional gauge theories. We emphasize that all of the ingredients necessary to construct the
Standard Model (or MSSM) are present in these compactifications, which are able to realize:
\begin{itemize}
	\item Non-abelian gauge symmetry with $G=U(N)$, $SO(2N)$, or $Sp(2N)$ for each brane.
	\item Chiral matter at brane intersections, with natural family replication.
	\item Hierarchical masses and mixing angles, dependent on geometry in the Calabi-Yau.
\end{itemize}
In type IIa, all of these effects are described by geometry, and one can often employ CFT techniques for their calculation. In the presence of orientifold planes, which are needed for globally consistent supersymmetric models, these theories can realize all of the representations present
in a standard Georgi-Glashow SU(5) grand unified theory (GUT). Such theories are often know
as D-braneworlds or type II orientifold compactifications\footnote{In this framework, the first globally consistent models with
chiral matter were presented in \cite{Angelantonj:2000hi,Blumenhagen:2000fp,Aldazabal:2000cn} and the first supersymmetric globally consistent
models with chiral matter were presented in \cite{Cvetic:2001tj,Cvetic:2001nr}. For comprehensive reviews and further information, see \cite{Blumenhagen:2005mu,Marchesano:2007de,Blumenhagen:2006ci} and references therein.}.

Despite their success in realizing gauge symmetry and chiral matter content, initial studies of D-braneworlds
did not give rise to some important parameters in particle physics. In particular, in \emph{all} weakly coupled
D-braneworlds, both the Majorana neutrino mass term $M_R \, N_R \, N_R$ and the Georgi-Glashow top-quark
Yukawa coupling
$10\, 10\, 5_H$ are forbidden in string perturbation theory by global $U(1)$ symmetries. 
In many concrete realizations these same global symmetries also forbid other Yukawa couplings, giving rise to massless families
of quarks or leptons, in direct contradiction with experiment.

In \cite{Blumenhagen:2006xt,Ibanez:2006da,Florea:2006si}, it was shown that euclidean D-brane instantons in D-braneworlds
can generate non-perturbative corrections to superpotential couplings involving charged matter. Thus, all of the mentioned
coupling issues can, in principle, be ameliorated by non-perturbative effects. The instanton corrections are exponentially suppressed,
with the factor depending on the volume of the cycle wrapped by the instanton in the Calabi-Yau, and therefore can account
for the large hierarchies seen in nature. Taking into account  D-instantons, then, the global $U(1)$ symmetries which forbid couplings
in perturbation theory are a virtue of these compactifications, rather than a drawback. For a comprehensive review of D-instantons in type II string theory, see \cite{Blumenhagen:2009qh} and references therein.

As there are comprehensive reviews of both
generic aspects of D-braneworlds \cite{Blumenhagen:2005mu,Marchesano:2007de,Blumenhagen:2006ci} and D-instantons \cite{Blumenhagen:2009qh}, we intend these notes to be a short
review of both topics. For the sake of brevity, we often omit in-depth derivations, choosing instead to give the reader an
intuition for the geometry of these models and the corresponding particle physics. We hope that they are sufficient to prepare
the reader to read either the existing literature or the in-depth reviews.

These lectures are organized as follows. In section \ref{sec:background} we give a rudimentary introduction to D-branes and explain 
how they
give rise to gauge theories. Based on scales in the theory, the possibility of large extra dimensions is discussed.
In section \ref{sec:CFT} we discuss important aspects of chiral matter. We begin by briefly reviewing conformal
field theory techniques and then use open string vertex operators to derive the supersymmetry condition as a function
of angles between branes. We introduce the notion of the orientifold projection, and discuss the appearance of particular representations
in this context. In section \ref{sec:global consistency} we discuss conditions required for global consistency of type
II orientifold compactifications. We derive the conditions on homology necessary for Ramond-Ramond tadpole
cancellation and present the generalized Green-Schwarz mechanism, as well as the constraints they impose on
chiral matter. In section \ref{sec:toroidal orbifold} we discuss the basics of toroidal orbifolds and present a Pati-Salam
model. In section \ref{sec:couplings} we discuss the appearance of Yukawa couplings in string perturbation theory via CFT techniques and present examples of two important couplings forbidden in perturbative theory. In section \ref{sec:instantons}, we present the basics of D-instantons. We discuss details of gauged axionic shift symmetries which are important for superpotential corrections. We also discuss details of charged and uncharged zero modes in terms of both CFT and sheaf cohomology, and present a concrete example of the instanton calculus. Finally, in section \ref{sec:quivers}
we discuss the advantages and disadvantages of the ``bottom-up" approach, which involves studying
quiver gauge theories rather than fully defined orientifold compactifications.

\section{Generic Background: D-branes, String Parameters, and Scales \label{sec:background}}
Early attempts to understand the physics of superstring theory involved the study of the Ramond-Neveu-Schwarz (RNS) action,
a (1+1)-dimensional superconformal worldsheet action with spacetime as the target space, which must be ten-dimensional to cancel
the worldsheet conformal anomaly. Rather than describing particle worldlines, the RNS action describes the physics
of string ``worldsheets" embedded in ten-dimensional spacetime. There are only two possibilities for the topology of the superstring:
either an $S^1$ or the interval, describing closed and open strings, respectively. The closed strings therefore have no
special points, but the open strings do, and it is important to ask what boundary conditions must be imposed at their
endpoints.

Very early in the history of string theory, it was realized that the open string equations of motion allow for two
types of boundary conditions, known as Neumann and Dirichlet. As an open string propagates in ten-dimensional spacetime,
it can have either type of boundary condition in each dimension of spacetime. An open string with $(p+1)$ Neumann dimensions and $(9-p)$ Dirichlet
dimensions has boundary conditions given by
\begin{align}
	\mu = 0,\dots, p \qquad \del_\sigma X^\mu|_{\sigma=0,\pi} = 0 \notag \\
	\mu = p+1, \dots, 9 \qquad \del_\tau X^\mu|_{\sigma=0,\pi}=0,
\end{align}
where $\sigma$ is the worldsheet coordinate along the string and $\tau$ is the worldline proper time in the particle
limit. A look at the definition of the Dirichlet condition might worry the reader, as the $\tau$-derivative vanishing means that the ends of the string
are ``stuck" at particular points in spacetime, which would seem to break Poincar\' e invariance. For this very reason early work on open
strings did not consider the possibility of Dirichlet boundary conditions.

It was realized, however, that there are issues with ignoring the possibility of Dirichlet boundary conditions. In particular,
it was soon realized that type IIa and type IIb superstring theory are T-dual to one another. The simplest statement of the duality is that type IIb
compactified on an $S^1$ of radius $R$ gives the same physics as type IIa compactified on an $S^1$ of radius $\frac{\alpha '}{R}$, 
where $\alpha '=l_s/(2\pi)^2$ depends on the string length $l_s$. A basic fact about the duality is that it exchanges Neumann and Dirichlet
boundary conditions of the open strings on the circular dimension, and thus if the duality
is to hold, Dirichlet and Neumann boundary conditions must be on the same footing.

The key insight which solved the issue about Poincar\' e invariance \cite{Polchinski:1995mt} was that open strings end on objects which themselves
carry energy, providing an object to which the momentum at the ends of the open string can escape. These objects, given the name D-branes based
on the importance of Dirichlet boundary conditions, source Ramond-Ramond charge from the closed string sector. A Dp-brane is an object on which an open string with $(p+1)$ Neumann dimensions and $(9-p)$ Dirichlet dimensions can end. The
endpoints of the open string can only move in the Neumann dimensions, and therefore the strings
are confined to the Dp-brane.

A massless open string which starts and ends on the same D-brane is interpreted as a gauge boson, since string quantization shows that it transforms in the
adjoint of some group $G$, usually $U(N)$, $SO(2N)$, or $Sp(2N)$. One interesting question is whether there is anything to be learned from the fact
that a Yang-Mills theory with gauge group $G$ is confined to some submanifold of the total spacetime, whereas the closed string (gravitational) sector
propagates in the full spacetime. Indeed, ignoring small constant factors for the time being\footnote{One can be more precise, of course, by expanding the DBI action to leading order in $\alpha '$ to
obtain the standard gauge kinetic term with appropriate factors. Here, we are just interested in consequences of dimensional analysis.},
the spacetime effective action $S$ contains a gauge term for the Dp-brane
and a gravitational term as
\begin{equation}
S\supset (\frac{M_s}{g_s})^{p-3} \int_{\bR^{3,1}\times\pi_a} F_{ab}F^{ab} + \frac{M_s^8}{g_s^2} \int_{\bR^{3,1}\times\cM} \cR_{10d},
\end{equation}
where the string mass $M_s=(\frac{1}{2\pi\alpha '})^{1/2}$ is the natural mass scale in the theory, $\pi_a$ is the $(p-3)$-cycle in the Calabi-Yau
manifold $\cM$ wrapped by the Dp-brane, and $g_s$ is the string coupling constant.
Dimensionally reducing to four dimensions, we obtain
\begin{equation}
S \supset (\frac{M_s}{g_s})^{p-3} \,\, \cV_{\pi_a} \int_{\bR^{3,1}} F_{\mu\nu}F^{\mu\nu} +\frac{M_s^8}{g_s^2} \,\, \cV_\cM  \int_{\bR^{3,1}} \cR_{4d}
\end{equation}
from which we can read off
$M_p^2 \sim \frac{M_s^8}{g_s^2} \cV_\cM$ and $\frac{1}{g^2_\text{YM}} \sim \frac{M_s^{p-3}}{g_s} \cV_{\pi_a}$. If the geometry is factorizable
such that $\cV_\cM = \cV_{\pi_a} \, \cV_t$, where $\cV_t$ is the volume of the dimensions in $\cM$ transverse to the Dp-brane,
it immediately follows that 
\begin{equation}
M_p^2 \,\, g^2_{YM} \sim \frac{M_s^{11-p}}{g_s} \, \cV_t.
\end{equation}
This relation between known parameters on the left-hand side and string-theoretic parameters on the right-hand side allows for the following important
observation: in the braneworld scenarios, a low string scale $M_s$ allows for the possibility of large extra dimensions transverse to the brane, whose
volume is $\cV_t$. The possibility of large extra dimensions was explored in \cite{ArkaniHamed:1998rs,Antoniadis:1998ig}, which looked in particular at the possibility
of two large extra dimensions.

Being a bit more precise, the dynamics of massless open
string modes in the worldvolume are given by the Dirac-Born-Infeld (DBI) action plus the Wess-Zumino (WZ) action.
\begin{align}
	S_\text{eff} &= S_\text{DBI} + S_\text{WZ} \notag
\end{align}
Together they form the relevant worldvolume Lagrangian to
leading order in the string coupling and derivatives. Looking
to the subset ${a,b}$ of
the ten-dimensional spacetime indices $M,N$ along the worldvolume
of the D-brane, the actions are given by\footnote{We realize we
are very brief, and refer the reader to \cite{Blumenhagen:2006ci} for more details.}
\begin{align}
\label{eqn:DBIWZ}
	S_\text{DBI} &= -\mu_p \int d^{p+1}x \,\, e^{-\phi} \,\, \sqrt{-det(G_{ab}+B_{ab}+2\pi \alpha ' F_{ab})} \notag \\
	S_\text{WZ} &= -\mu_p \int d^{p+1}x \,\,\, \text{tr} \, e^{2\pi \alpha ' \cF} \wedge \sqrt{\frac{\hat A (\cR_T)}{\hat A (\cR_N)}} \wedge \bigoplus_q C_q
\end{align}
where the Ramond-Ramond charge of a $p$-brane is $\mu_p = 2\pi l_s^{-p-1} = (2\pi)^{-p} (\alpha ')^{(-p-1)/2}$
and the fields are defined $G_{ab} \equiv \del_a X^M \del_b X^N g_{MN}$ and $B_{ab} = \del_a X^M \del_b X^N B_{MN}$. $\hat A(x)$ is the A-roof genus, given in terms of the Pontryagin classes $p_i$ of a real bundle $x$ as  $\hat A(x) = 1 - \frac{1}{24} p_1 + \frac{1}{5760}(7p_1^2-4p_2) + \dots$, and $\cR_T$ and $\cR_N$ are the curvature forms of the tangent bundle and normal bundle of the brane worldvolume, respectively.

The key
physics to note is that the DBI action describes the coupling
of the open string gauge field modes in $F$ to the massless NSNS sector, that is, the dilaton, metric and two-form. The WZ
action, on the other hand, describes the coupling of the Ramond-Ramond forms $C_q$ which charge the D-branes to the gauge field
$F$. As we will see in section \ref{sec:global consistency}, the WZ
action is a useful tool for deriving and understanding global
consistency conditions of type II orientifold compactifications.

\section{Massless Spectrum and Conformal Field Theory \label{sec:CFT}}
Having introduced D-branes and the fact that gauge theories are confined to them in section \ref{sec:background},
in this section we present details related to those gauge theories. Specifically, using conformal field theory techniques
we will discuss the appearance of gauge bosons and chiral matter, as well as details related to the presence of
orientifolds, whose presence often allows for more interesting four-dimensional particle physics.

A particularly natural arena in which to discuss the massless spectrum of D-braneworlds is that
of conformal field theory, where quantization techniques allow us to directly identify interesting
properties of massless superstrings. For brevity, we present only the details relevant for our presentation,
and refer the interested reader to \cite{FMS1, FMS2, Knizhnik, DFMS, Polchinski, PeskinTASI} and references therein for more details on the BRST
quantization of superstrings.

We consider open strings attached to the D-brane, with Dirichlet boundary conditions transverse to the D-brane 
and Neumann boundary conditions along the D-brane worldvolume. In the conformal field theory description, we have
a concrete representation of massless states, as they are given by vertex operators. The worldsheet
fermions appearing in vertex operators have two possibilities for the boundary conditions,
Neveu-Schwarz and Ramond, differentiated by a sign in $\psi$ upon going around the spatial
direction of the closed superstring, with the Ramond sector carrying the minus sign.

\subsection{Non-abelian Gauge Symmetry}

In the NS sector with superconformal ghost number $\phi=0$, the vertex operator for the gauge boson is given by
\begin{equation}
V_{A_\mu}=\xi_\mu \partial_zX^\mu e^{ik \cdot X},
\end{equation}
where $\mu \in \{0,1,2,3\}$ and $\xi_\mu$ is the polarization vector in the target space, making this a spin 1 field. Here we employ
radial quantization by mapping the Euclidean
worldsheet coordinates $(\tau_E,\sigma)$ to complex coordinates $z,\ov z$ as
\begin{eqnarray}
z=e^{\tau_E+i\sigma}\nonumber\\
\bar{z}=e^{\tau_E-i\sigma}.
\end{eqnarray}
Since these are open strings, there is a worldsheet boundary condition $0\le\sigma\le\pi$, which is equivalent
to $\im z\ge 0$. Instead of considering two sets of Virasoro generators defined on the upper-half plane $\im z\ge 0$,
we employ the doubling trick\footnote{The conformal field theory references should contain more details, for the
reader unfamiliar with the doubling trick.} in order to consider one set of Virasoro generators on the \emph{whole}
complex plane with holomorphic coordinate $z$.

The NS vertex operator in the (-1) ghost pictures is given by:
\begin{eqnarray}
\label{eqn:vert gauge boson}
V_{A_I}(-1)=\xi_\mu e^{-\phi} \psi^\mu e^{ik \cdot X},
\end{eqnarray}
where the conformal dimensions of the fields appearing in the vertex operator are
\begin{eqnarray}
\left[ e^{\alpha \Phi} \right]=-\frac{\alpha (\alpha +2)}{2} \qquad \qquad
\left[\psi \right]=\frac{1}{2} \qquad \qquad [e^{i\, k\cdot X}] = \frac{\alpha' k^2}{2}.
\end{eqnarray}
Calculating the conformal dimension of \eqref{eqn:vert gauge boson}, we obtain 
$\left[V_{A_I}\right]=\frac{1}{2} + \frac{1}{2} + \frac{\alpha' k^2}{2}$. The requirement
that this be equal to one shows that this vertex operator corresponds to a massless field, in particular a massless spin-1 field confined to the worldvolume of the D-brane. It therefore has
a natural interpretation as a gauge boson.

Thus far, the vertex operators discussed correspond to the massless degrees of freedom in a pure $U(1)$ gauge theory, living
on the D-branes. As is well known from the theory of open strings, there is a generalization which corresponds to adding
degrees of freedom at the endpoints of the strings. These degrees of freedom are allowed because they break no symmetries (conformal,
Poincar\' e, etc) of the worldsheet theory and are known as Chan-Paton factors. Though this is a trivial generalization of the
worldsheet theory, it has profound implications for spacetime physics. In particular, for $a\in\{1, \dots, N\}$, the Chan-Paton
factor $\Lambda_a$ associated with one end of an open string corresponds to that end being confined to a stack of $N$ coincident
D-branes. Adding a Chan-Paton factor for each end of the massless open string associated with \eqref{eqn:vert gauge boson},
the generalized vertex operator is
\begin{equation}
\label{eqn:vert gauge boson w chan-paton}
V_{A_\mu}=\xi _\mu e^{-\phi}\psi ^\mu e^{ik \cdot X}(\Lambda _a\otimes \bar{\Lambda}_b)
\end{equation}
with $a,b\in\{1,\cdots N\}$, $\Lambda_a$ the fundamental, and $\ov \Lambda_b$ the antifundamental. In the absence of orientifolds, the $N^2$ degrees of freedom in the Chan-Paton factors transform
in the adjoint of $U(N)$. Thus, a stack of $N$ coincident D-branes has a $U(N)$ gauge theory living on its worldvolume.

\subsection{Orientifold Projection}

In addition to D-branes, type II string theory also allows for the presence of another type of
object which carries Ramond-Ramond charge. These objects have a fixed negative tension and are known as orientifold planes,\FIGURE{
\centering
 \includegraphics[scale=.6]{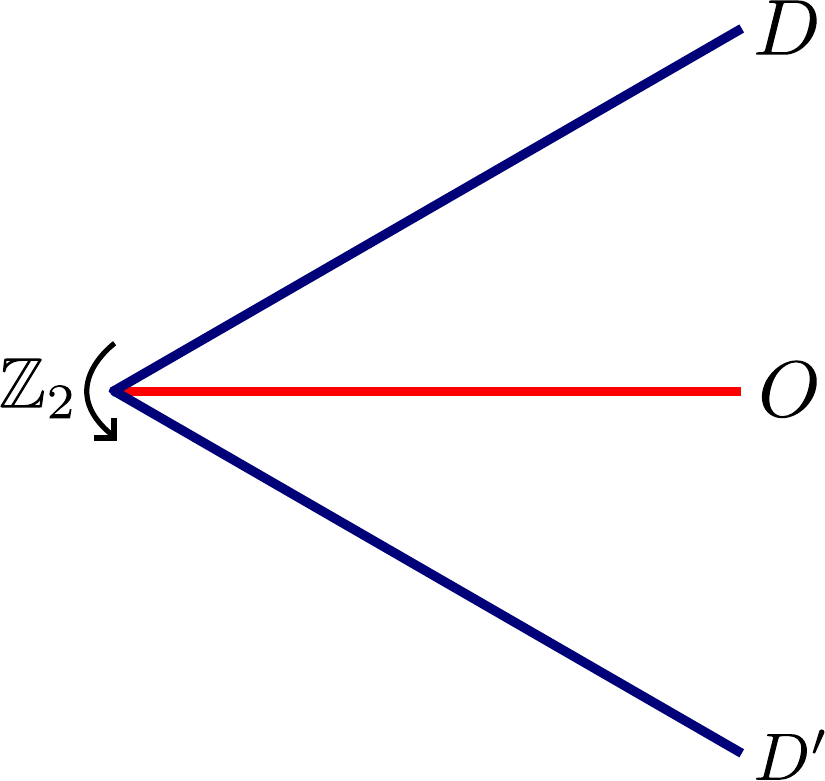}
 \caption{Graphical depiction of a brane $D$ and its orientifold image $D'$, as well the the orientifold $O$, which is the fixed point locus of the specified $\bZ_2$ action.}
 \label{fig:dod}
}
\noindent and it is this negative tension which allows for supersymmetric globally consistent models, as we will see in section \ref{sec:global consistency}.
To specify an orientifold compactifications in type II, one must provide an orientifold projection, in addition to a Calabi-Yau
manifold. The orientifold projection is a combination of three actions:
\begin{itemize}
	\item[$\bullet$] $\Omega: \sigma \mapsto -\sigma$, the worldsheet parity operator
	\item[$\bullet$] $(-)^F$, an action on worldsheet fermions
	\item[$\bullet$] a $\bZ_2$ involution on the Calabi-Yau.
\end{itemize}
The $\bZ_2$ involution, which must be antiholomorphic in type IIa and holomorphic in type IIb, generically has a non-trivial
fixed point locus, where the orientifold planes sit. 

The involution acts on non-trivial cycles in the Calabi-Yau,
and therefore it also acts on any D-brane wrapping a non-trivial cycle. 
Associated to any D-brane in an orientifold compactification, therefore,
is an image brane, as depicted in Figure \ref{fig:dod}. The $\bZ_2$ involution fixes the homology of the O-planes, which via global consistency
has profound implications for particle physics. We postpone this detailed discussion until section \ref{sec:global consistency}.

In addition to implications for the homology of D-branes and O-planes, the orientifold projection imposes constraints on the physical
states of the theory. If the cycle wrapped by a stack of $N$ D-branes is not invariant under the orientifold, then there are no additional
conditions on the physical state, and both that stack and its image stack give rise to $U(N)$ gauge symmetry. If the cycle is orientifold
invariant, one must distinguish between two cases:
\begin{itemize}
	\item[$\bullet$] the cycle is pointwise fixed
	\item[$\bullet$] the cycle is fixed, but only in homology.
\end{itemize}

In both cases the orientifold projection imposes a constraint on the Chan-Paton factors, given by $(\Lambda _a\otimes 
\bar{\Lambda}_b)=\pm(\Lambda _a\otimes \bar{\Lambda}_b)^T$, where the different signs are for the different cases in
homology. The extra constraint changes the gauge theory on the D-brane from $U(N)$ to $SO(2N)$ or $Sp(2N)$, depending on
sign, where the details are dependent upon whether one is in type IIa or type IIb. We refer the reader to section
2.2.8 of  \cite{Blumenhagen:2006ci} for more details.

\subsection{Chiral Matter and Representations\label{sec:chiral matter}}
Now that we have introduced the very basics of orientifolds, we have the requisite background for describing generic matter content
in weakly coupled type II string theory. The basic idea is that an open string could, in general, end on two different stacks
of D-branes, $a$ and $b$. We say that such a string is in the $ab$-sector, and therefore the adjoint representation of the
$a$-stack and $b$-stack correspond to the $aa$-sector and the $bb$-sector, respectively. Orientifolds allow for further generality,
as now each D-brane has an image brane, allowing for the extra possibilities of strings in the $aa'$-sector, $ab'$-sector,
and $bb'$-sector.

\FIGURE{
 \centering
 \includegraphics[scale=.4]{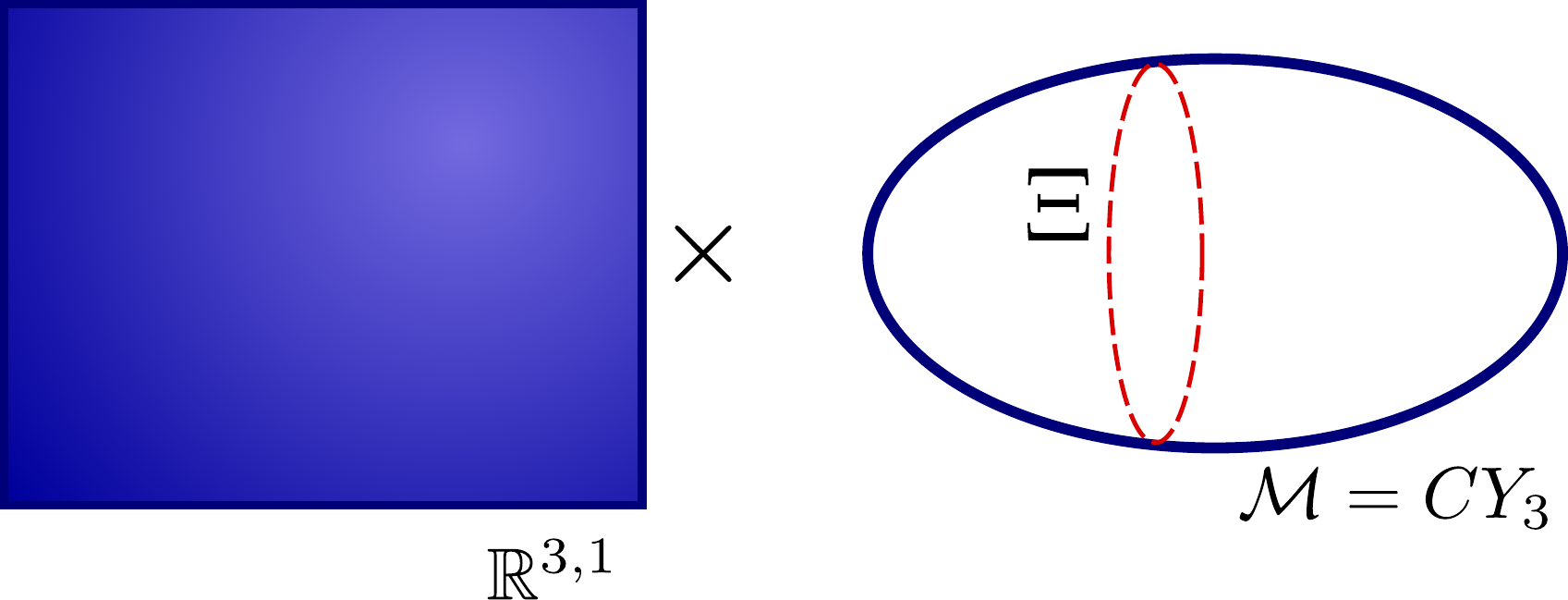}
 \caption{\small}
 \label{fig:dbrane cycle}
 \caption{A target space view of a spacetime filling gauge D-brane, which wraps a non-trivial cycle in the Calabi-Yau.}
}

A natural question to ask is whether there is a geometric way to see the appearance of matter from the Higgsing of the
adjoint representation of some higher gauge group. Considering a stack of $(N+M)$ D-branes on a generic cycle, called $a$,
the only brane sector is the $aa$-sector, which corresponds to the adjoint of $U(N+M)$ and has $(N+M)^2=N^2+M^2+2NM$ 
degrees of freedom. Performing a geometric deformation where one unfolds the stack of $(N+M)$ D-branes into stacks of $N$ and $M$ D-branes,
called $b$ and $c$, the degrees of freedom from the $aa$-sector now lie in the $bb$, $cc$, or the $bc$ sector. These sectors respectively give rise
to the adjoints of $U(N)$ and $U(M)$, as well as bifundamentals $(N,\ov M)$ and $(\ov N, M)$. This is a concrete geometric
interpretation in string theory for how the adjoint of $U(N+M)$ breaks to the adjoint of $U(N)$, $U(M)$ and bifundamentals. In this simple
example, the matter fields in the bifundamental representations live at the intersection of
two branes.

Having motivated the notion that matter lives at the intersection of different stacks of D-branes, we focus on intersecting
D6-branes in type IIa for a concrete discussion of the quantization of an open string between two stacks of intersecting branes.
Calling the three-cycles wrapped by two stacks of D6-branes $\pi_a,\pi_b\in H_3(\cM,\bZ)$, we write
\begin{align}
\pi_a=N_{aI}A^I+M_a^JB_J \notag \\
\pi_b=N_{bI}A^I+M_b^JB_J,
\end{align}
where we have used the symplectic basis of three-cycles $A^I$ and $B_I$. They satisfy
\begin{eqnarray}
A^I\circ B_J=\delta ^I_J, \qquad 
\int _{A^I}\alpha_J = \delta _{IJ}, \qquad
\int _{B_I}\beta_J =- \delta ^{IJ}
\end{eqnarray}
 where $\alpha_J$ and $\beta^J$ are the dual basis of three-forms on the Calabi-Yau.
 From this it is straightforward to calculate the topological intersection number
 \begin{eqnarray}
 \pi _a \circ \pi _b= N_{aI}M_b^I-M_a^IN_{bI}.
\end{eqnarray}
This already has interesting implications for physics: if, for example, $\pi_a\circ \pi_b=3$, 
then these D-branes intersect at three points in the Calabi-Yau, giving three copies of
whatever quantized matter lives at a single intersection. This is quite naturally interpreted
as family replication.

The quantization of open strings living at the intersection of two D-branes\footnote{We recommend appendix A of
\cite{Cvetic:2006iz} for all the details.} utilizes the
standard open string quantization, together with the non-trivial boundary conditions
\begin{eqnarray}
\label{eqn:boundary conditions}
\partial _\sigma X^{2I-1}=X^{2I}=0 \quad \text{at} \quad \sigma =0 \nonumber \\
\partial _\sigma X^{2I-1}+(\tan  {\pi \theta _I})\partial _\sigma X^{2I}=0 \quad \text{at} \quad \sigma =\pi \nonumber\\
X^{2I}-(\tan  {\pi \theta _I}) X^{2I-1}=0 \quad \text{at} \quad \sigma =\pi \nonumber
\end{eqnarray}
where $I=\in\{2,3,4\}$\footnote{Here we depart from conventions elsewhere in the literature, which often
use $\theta_1$, $\theta_2$, and $\theta_3$ for the angles in each of the three complex internal
dimensions. We use the labeling $0,\dots,9$ for real dimensions and $0,\dots,4$ for complexified
dimensions, thus $2,3,4$ for the angles between branes on the internal space.}. 
Going to complexified notation for the worldsheet bosons and quantizing, with boundary conditions taken into account, 
the expansion in terms of oscillators is given by
\begin{equation}
Z^I=X^{2I-1}+iX^{2I}=\sum _{n \epsilon {\bZ}}\frac{\alpha _{n-\theta_I}^I}{n-\theta_I}z^{-n+\theta_I}+
\sum _{n \epsilon {\bZ}}\frac{\tilde{\alpha} _{n+\theta_I}^I}{n+\theta_I}\bar{z}^{-n-\theta_I},
\end{equation}
and we note that since $\tilde{\alpha}_{n+\theta_I}^{I\dagger}=\alpha_{n+\theta_I}^I$, $Z\mapsto \ov Z$ under $\theta _I \mapsto -\theta_I$ . 
The only non-vanishing commutator is $[\alpha^I_{n\pm\theta}, \alpha^{I'}_{m\mp \theta}] = \pm \, m \,\delta_{n+m} \delta^{II'}.$ We can also complexify worldsheet fermions, giving
\begin{align}
\text{Neveu-Schwarz Sector:}& \qquad \Psi ^I=\psi ^{2I-1}+i\psi ^{2I}=\sum _{r\in\bZ+\frac{1}{2}}\psi_{r-\theta_I}z^{-r-\frac{1}{2}+\theta _I} \notag \\
\text{Ramond Sector:}& \qquad \Psi ^I=\psi ^{2I-1}+i\psi ^{2I}=\sum _{r\in\bZ}\psi_{r-\theta_I}z^{-r-\frac{1}{2}+\theta _I},
\end{align}
where the only non-vanishing anticommutator is $\{\psi^I_{m-\theta_I},\psi^{I'}_{n+\theta_I}\} = -\delta_{mn} \, \delta^{I,I'}$. We refer the reader to
\cite{Cvetic:2006iz} for more details on oscillator quantization for D-branes at intersecting angles, including zero point energies and mass formulae.


Instead, we show the equivalent physics using the vertex operator formalism of CFT. In the vertex operator formalism,
it is the presence of bosonic twist fields $\sigma_\theta$ \cite{DFMS} that ensure the boundary conditions \eqref{eqn:boundary conditions}
for D6-branes intersecting at non-trivial angles $\theta_I$. As one might expect, we will have vertex operators for both fermions and bosons
living at the intersections of two branes. From studying their conformal dimensions, we will extract mass formulae and show that, though the
fermion is always massless, the mass of the boson depends on the angles of intersection.

Recall that in quantizing the superstring, one often chooses to bosonize the worldsheet fermions $\Psi^M$ with
$M\in \{0,\cdots,9\}$, rather than working with the fermions directly. Usually each of the five complexified worldsheet fermions
is bosonized as
\begin{align}
\text{Neveu-Schwarz Sector:}& \qquad \Psi^M \cong e^{iH_M} \notag \\
\text{Ramond Sector:}& \qquad \Psi^M \cong e^{i(1\pm\frac{1}{2})H_M},
\end{align}
where the half integer in the last line is present in order to take care of the Ramond boundary conditions on
the worldsheet. The $\pm$ ambiguity corresponds to each complexified worldsheet fermion having spin
$\pm\frac{1}{2}$, and the $2^5$ sign choices reflect that fact that the Ramond sector ground state is a $32$-dimensional
Dirac spinor in ten dimensions.

The key difference between a standard open superstring and an open superstring at the intersection of two
D-branes is the boundary conditions, which must be taken into account. As they only apply in the internal
dimensions, they only change three of the complexified worldsheet fermions, which become
\begin{align}
\text{Neveu-Schwarz Sector:}& \qquad \Psi^I \cong e^{i\theta_I H_I} \notag \\
\text{Ramond Sector:}& \qquad \Psi^I \cong e^{i(\theta_I\pm\frac{1}{2})H_I},
\end{align}
with $I\in\{2,3,4\}$, where the sign in the last line depends crucially on how the angles are defined. Here
we choose the conventions $0<\theta_I<1$ for $I=2,3$ and $-1<\theta_4\leq 0$.
The two complexified worldsheet fermions which are not subject to boundary conditions form a two-component Weyl
spinor in four dimensions, which we write as $S^\alpha$ in the vertex operators.

Having discussed the relevant ingredients, we would like to explicitly write 
two vertex operators, in the NS-sector and R-sector, for open strings stretched
between spacetime filling D6-branes with non-trivial intersection in the Calabi-Yau.
Omitting Chan-Paton factors, since they aren't immediately relevant to the discussion, 
the vertex operators are
\begin{eqnarray}
\label{eqn:int vertex op}
V_{-1}=e^{-\phi}\prod _{I=2}^3 \sigma_{\theta_I}e^{i\theta_I H_I}\sigma_{1+\theta_4}e^{i(1+\theta_4)H_4}e^{ik\cdot X}\nonumber\\
V_{-\frac{1}{2}}=u_\alpha e^{-\frac{\phi}{2}}S^\alpha \prod _{I=2}^3 \sigma_{\theta_I}e^{i(\theta_I-\frac{1}{2}) 
H_I}\sigma_{1+\theta_4}e^{i(\frac{1}{2}+\theta_4)H_4}e^{ik\cdot X} ,
\end{eqnarray}
respectively. To calculate the mass of these states, one must know the conformal weights of the fields appearing
in the vertex operators, which are given by
\begin{align}
[e^{\alpha\phi}]=-\frac{\alpha(\alpha+2)}{2}\qquad \qquad [\sigma_\theta]=\frac{\theta(1-\theta)}{2}\qquad \qquad [e^{iaH_I}] = \frac{a^2}{2} \notag \\ \notag \\
[e^{i k\cdot X}] = \frac{\alpha ' k^2}{2} \qquad \qquad [S^\alpha] = [e^{i(\pm\frac{1}{2}H_1 \pm\frac{1}{2} H_2)}] = 2\frac{(\pm\frac{1}{2})^2}{2} = \frac{1}{4}.
\end{align}
Knowing these, it is straightforward to calculate the conformal weights of the vertex operators \eqref{eqn:int vertex op} to be
\begin{eqnarray}
\left[V_{-1}\right]=\frac{1}{2}+\sum_{I=2}^3\left(\frac{\theta_I(1-\theta_I)}{2}+\frac{1}{2}\theta_I^2\right) 
+\frac{(\theta_4 +1)(-\theta_4)}{2}+\frac{(\theta_4 +1)^2}{2}\nonumber\\
=\frac{1}{2}+\frac{1}{2}\sum _{I=1}^3 \theta_I +\frac{1}
{2}+\frac{\alpha ' k^2}{2} \\ \nonumber \\ \nonumber \\
\left[V_{-\frac{1}{2}}\right]=\frac{3}{8}+\frac{1}{4}+3\frac{1}{8}+\frac{\alpha ' k^2}{2}.
\end{eqnarray}
The mass formulae \cite{DFMS, FMS1, FMS2, Berkooz:1996km,Bachas:1995ik} are derived from the requirement that the conformal dimension be one, yielding
\begin{align}
	\alpha ' \, m^2 = \sum_{I=1}^3 \theta_I \qquad \qquad \text{and} \qquad \qquad \alpha ' m^2 = 0,
\end{align}
for the spacetime bosons and fermions, respectively. Therefore, if the sum of the three angles is negative, zero, or positive
then the boson is tachyonic, massless, and massive, respectively. In the case where the boson is massless, the NS-sector boson $V_{-1}$
becomes massless and forms a supermultiplet with the R-sector fermion $V_{-\frac{1}{2}}$. Thus, $\sum_I \theta_I=0$ is the local
condition for intersecting branes to give rise to supersymmetric matter.

The angle condition is a local picture of mutually supersymmetric branes. The global picture is that for a
D-brane to give rise to a supersymmetric gauge theory, it must
wrap a supersymmetric cycle, given by a special Lagrangian
in type IIa or a holomorphic divisor in type IIb. For the
effective four-dimensional theory to be supersymmetric, the
branes must preserve the \emph{same} supersymmetry. It has been shown that the global conditions
for two D6-branes on special Lagrangians to preserve the
same supersymmetry reduces locally to the condition on angles.

\TABLE{
\centering \vspace{3mm}
\label{table:spectrum}
\begin{tabular}{|c|c|}
\hline
Representation  & Multiplicity \\
\hline $\Yasymm_a$
 & ${1\over 2}\left(\pi_a\circ \pi'_a+ \pi_a\circ \pi_{{\rm O}6}
\right)$  \\
 $\Ysymm_a$
      & ${1\over 2}\left(\pi_a\circ \pi'_a-\pi_a\circ \pi_{{\rm O}6}
\right)$   \\
      $(\fund_a,\antifund_b)$
       & $\pi_a\circ \pi_{b}$   \\
        $(\fund_a,\fund_b)$
	 & $\pi_a\circ \pi'_{b}$
	 \\
	 \hline
	 \end{tabular}
	 \caption{Representations and multiplicities for chiral matter at the intersection of two D6-branes.} 
}

The vertex operators \eqref{eqn:int vertex op} also generically include Chan-Paton factors, which might satisfy further constraints
due to the orientifold projection. The generic case we would like to discuss is the structure of Chan-Paton factors at the intersection of 
two gauge D-branes, in which case the factors are a tensor product of some combination of fundamentals and antifundamentals. That is, for
gauge branes with $U(N_a)$ and $U(N_b)$ gauge symmetry, the possibilities are $(\fund_a, \fund_b)$, $(\fund_a, \antifund_b)$, $(\antifund_a, \fund_b)$
and $(\antifund_a, \antifund_b)$, where the choice between fundamental and antifundamental depends on the direction of the string and whether or
not the string ends on a brane or its orientifold image. Thus, the most common possibility is that chiral matter appearing at the intersection
of two D-branes is in the bifundamental representation. The possible representations and multiplicities of chiral matter are listed in Table \ref{table:spectrum}.

\FIGURE{
 \centering
 \includegraphics[scale=.8]{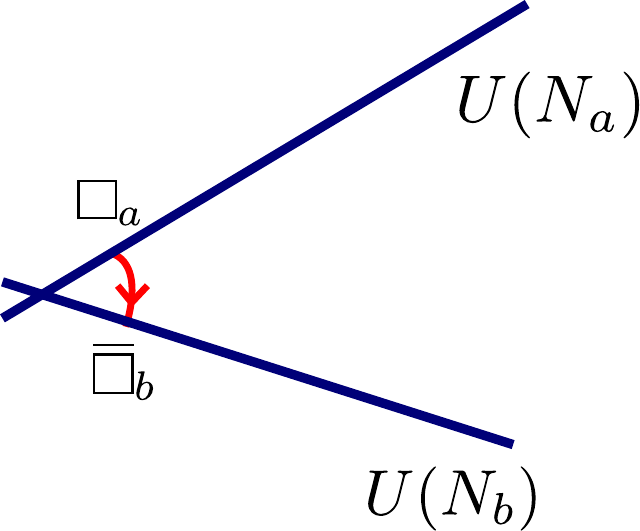}
 \caption{A bifundamental string beginning on a $U(N_a)$ brane and ending on a $U(N_b)$ brane, with appropriate Chan-Paton
 factors at the endpoints.}
 \label{fig:upyukawa}
}

There are two special cases which are interesting to discuss, one of which arises in the table. First, we revisit the possibility that a string begins and ends on the same
$U(N_a)$ brane. In such a case, the Chan-Paton factors take the form $\fund_a \otimes \antifund_a = \text{Adj}_a \oplus 1$. That is, due to
the decomposition into a direct sum, the string beginning and ending on the same brane can transform in the adjoint representation, and
is therefore a gauge boson. Second, one might wonder about the properties of a string beginning on a brane and ending on its orientifold
image. In such a case the Chan-Paton factors take the form $\fund_a \otimes \fund_a = \Ysymm_a \oplus \Yasymm_a$, and therefore chiral matter can be in the
symmetric or antisymmetric representation of the gauge group. This is of great importance, for example, in $SU(5)$ GUT models. There the $10$
representation is needed, which can be realized as $10=\Yasymm_5$ in type II D-braneworlds.

Thus, we see that D-braneworlds can give rise to all of the ingredients necessary for realizing the gauge symmetry and matter content of the Standard Model. In particular, a stack of multiple D-branes give rise to non-abelian gauge symmetry, with the possibility of chiral matter living at the intersection of two D-branes. We have shown the presence of chiral supermultiplets locally at the intersection of two D-branes. Upon taking into account global aspects, namely the
need for compactification, the fact that branes can intersect multiple times in the internal space gives a geometric reason for family replication.

\section{Global Consistency \label{sec:global consistency}}
Since string theory gives rise to low energy gauge theories in a variety of ways, it has been important
throughout its history to address whether or not string theory gives rise to \emph{consistent}
gauge theories. For instance, the first superstring revolution was sparked in \cite{GreenSchwarz} when Green and Schwarz showed that the type I string in ten dimensions with $SO(32)$ gauge symmetry
is anomaly free. Over time, anomaly cancellation has been shown to arise naturally in many corners
of the landscape. In all corners, the conclusion thus far has been the same: the natural ingredients arising in
a string theory ensure the consistency of the low energy effective theory.

In this section, we will present known results for how this occurs in type II orientifold compactifications.
We begin with tadpole cancellation, which amounts to conditions on the homology of spacetime filling 
D-branes and O-planes that ensures the necessary cancellation of Ramond-Ramond charge on the internal space.
These global conditions on homology impose constraints on chiral matter which are \emph{necessary} for tadpole
cancellation. We will show that a subset of these constraints on chiral matter are precisely the conditions
for the cancellation of non-abelian anomalies. We will also show that the presence of Chern-Simons
couplings of Ramond-Ramond forms to $U(1)$ field strengths gives rise to a generalized Green-Schwarz mechanism
which cancels abelian and mixed anomalies. Couplings of this type also generically give a Stuckelberg mass
to the corresponding $U(1)$ gauge bosons, whose corresponding global symmetries impose phenomenologically
important selection rules on superpotential couplings.

\subsection{Ramond-Ramond Tadpole Cancellation}
Historically, by studying amplitudes arising in CFT descriptions, it was shown that certain one-loop cylinder,
Mobius strip, and Klein-bottle diagrams have infrared divergences due to the presence of massless
Ramond-Ramond tadpoles, which are required to cancel for consistency of the theory. With the advent of D-branes,
a geometric picture of tadpole cancellation became clear in the works of \cite{Aldazabal:2000dg,Blumenhagen:2002wn,Blumenhagen:2002vp}. Following those works and working
in type IIa, we examine the RR seven-form kinetic term of the ten-dimensional supergravity Lagrangian, along with relevant Wess-Zumino terms \eqref{eqn:DBIWZ} of the D-brane effective action
\begin{align}
S \supset -\frac{1}{4\kappa^2}\int_{\bR^{3,1}\times \cM} H_8 \wedge *H_8 \,\,\,+ \,\,\, \mu_6 \sum_a N_a \int_{\bR^{3,1}\times \pi_a} C_7 \,\,\, \notag \\ 
+ \,\,\, \mu_6 \sum_a N_a \int_{\bR^{3,1}\times \pi_a^{'}} C_7 \,\,\, - \,\,\, 4\mu_6 \sum_a N_a \int_{\bR^{3,1}\times \pi_{O6}} C_7,
\end{align}
where $H_8=dC_7$ is the field strength of the Ramond-Ramond seven-form which couples to D6-branes and O6-planes, $\pi_a$ is the three-cycle
wrapped by a D6-brane and $\pi_a '$ is wrapped by its orientifold image, and $\pi_{O6}$ is the three-cycle wrapped by the O6-plane. The
ten-dimensional gravitation coupling is $\kappa^2=\frac{1}{2}(2\pi)^7(\alpha ')^4$.
Given this action and the Poincar\' e dual $\delta(\pi_a)$ of $\pi_a$, the equation of motion is
\begin{equation}
\frac{1}{\kappa^2} \,d(*H_8) = \mu_6 \sum_a N_a \, (\delta(\pi_a) + \delta(\pi_{a^{'}})) \,\,\,-\,\,\, 4\mu_6 \, \delta(\pi_{O6})
\end{equation}
and we see from the left-hand side that the right-hand side is an exact form, and is therefore trivial in $H^3(\cM,\bZ)$. But Poincar\' e duality is an \emph{isomorphism} between cohomology and homology, and therefore the Poincar\' e dual of the right hand side must be trivial in homology, yielding
\begin{equation}
\label{eqn:tadpole}
\sum_a N_a \, ([\pi_a] + [\pi_{a^{'}}]) = 4 \, [\pi_{O6}],
\end{equation}
where $[\pi_a]$ is the homology class of the three-cycle $\pi_a$.

This is the D6-brane tadpole cancellation condition in type IIa orientifold compactifications. It is a condition on the homology of the cycles which the D6-branes and O6-planes wrap. Qualitatively, satisfying this condition ensures that the Ramond-Ramond charge is canceled on the internal space, which is necessary since the spacetime filling D6-branes and O6-planes source Ramond-Ramond charge and the directions transverse to them are a submanifold of the compact Calabi-Yau $\cM$. That is, the condition must be satisfied, as otherwise the flux lines would have nowhere to go in a compact manifold.

The condition on homology is the necessary and sufficient for cancellation of homological RR tadpoles, but the condition on homology induces
constraints on chiral matter which are necessary for tadpole cancellation. These are interesting in their own right. Using Table \ref{table:spectrum} 
and intersecting
\eqref{eqn:tadpole}\footnote{With indices switched. That is, the sum is over $b$, rather than $a$.} with another three-cycle $\pi_a$ wrapped by a D6-brane (in the case where orientifolds are absent), we obtain
\begin{equation}
	0 = \pi_a \circ \sum_b N_b \,\,\, \pi_b = \sum_b N_b \,\,\, I_{ab} = \sum_b N_b (\#(a,\ov b) - \#(\ov a,b)).
\end{equation}
This can be rewritten as $\#a = \#\ov a$, which is precisely the condition for non-abelian anomaly cancellation if $N_a>2$.
Generalizing to the case with orientifolds and image branes, the full condition is
\begin{align}
	\label{eqn:chiral tadpole constraint}
	N_a \ge 2&: \qquad \# a - \# \ov a + (N_a+4)\,\, \# \, \Ysymm_a + (N_a-4) \,\, \# \, \Yasymm_a = 0 \notag \\ \notag \\
	N_a = 1&: \qquad \# a - \# \ov a + (N_a+4)\,\, \# \, \Ysymm_a = 0 \,\,\, \text{mod} \,\,\, 3,
\end{align}
where the mod 3 condition for the $N_a=1$ case comes from the fact that there is no antisymmetric representation of a $U(1)$.
For more details of this derivation, we refer the reader to \cite{Cvetic:2009yh}.

Thus, we see that type II string theory provides a beautiful geometric picture for the existence of anomaly cancellation:
the Ramond-Ramond charge of spacetime filling branes must be canceled on the internal space, which yields a condition on
homology\footnote{It is important to note that, in addition to the necessary cancellation of homological Ramond-Ramond charge, one must also cancel K-theory charges, due to the fact that D-branes are classified by K-theory groups \cite{Witten:1998cd} rather than homology groups. We refer the interested reader to \cite{Blumenhagen:2005mu,Marchesano:2007de,Blumenhagen:2006ci} for more details.} that induces necessary constraints on chiral matter. 
These constraints on chiral matter happen to include the cancellation of non-abelian anomalies, but also include some genuinely stringy 
constraints.

\subsection{Generalized Green-Schwarz Mechanism \label{sec:Generalized Green-Schwarz Mechanism}}
While the constraints on chiral matter corresponding to the cancellation of non-abelian anomalies are satisfied
immediately if the homological tadpole cancellation condition is satisfied, tadpole cancellation does not cancel
the abelian, mixed abelian-non-abelian, and mixed abelian-gravitational anomalies. In \cite{Aldazabal:2000dg}
it was shown that there is a generalization of the Green-Schwarz mechanism \cite{GreenSchwarz} to the case of intersecting branes. 
The mechanism cancels the anomalies \cite{Aldazabal:2000dg,Blumenhagen:2002wn} by the gauging of axionic shift 
symmetries associated with the Ramond-Ramond forms.

We now address some of the details. Again for concreteness we work with the type IIa supergravity action. 
Expanding the exponential of the field strengths of the gauge fields in the Wess-Zumino action \eqref{eqn:DBIWZ}, each stack of $D6$-branes,
indexed by $a$, has Chern-Simons couplings of the form
\begin{equation}
\label{eqn:chernsimons}
\int_{\bR^{3,1}\times \pi_a} C_3 \wedge \text{Tr}(F_a \wedge F_a), \qquad \qquad \int_{\bR^{3,1}\times \pi_a} C_5 \wedge \text{Tr}(F_a),
\end{equation}
where $F_a$ is the gauge field strength on the brane.
As we are concerned with mixed anomaly cancellation in the effective four-dimensional gauge theory, we expand the Ramond-Ramond forms
in basis $(\beta^I,\alpha_J)$ Poincar\' e dual to the integral basis of three-cycles $(A^I,B_J)$ defined in section \ref{sec:chiral matter} as
\begin{equation}
C_3 = \Upsilon^I\,\alpha_I + \tilde \Upsilon_I \, \beta^I \qquad \text{and} \qquad C_5 = \tilde \Delta^I \, \alpha_I + \Delta_I \, \beta^I
\end{equation}
where the coefficients of $\alpha$ and $\beta$ are the four-dimensional axions and two-forms
\begin{align}
\Upsilon^I &= \int_{A^I} C_3 \qquad \qquad \tilde \Upsilon_{I} = -\int_{B_I} C_3 \notag \\
\tilde \Delta^I &= \int_{A_I} C_5 \qquad \qquad \Delta_{I} = - \int_{B^I} C_5.
\end{align}
Upon dimensional reduction of $\cN_a$ D6-branes on $\pi_a = Na_I\, A^I + M_a^I \, B_I$ , the generic Chern-Simons couplings \eqref{eqn:chernsimons} can be written
in terms of axionic couplings of the form
\begin{equation}
\label{eqn:4d terms}
\cN_a\,\,\int_{\bR^{3,1}} (N_{aI} \Upsilon^I - M_a^I \tilde \Upsilon_I) \wedge \text{Tr}(F_a \wedge F_a),\qquad \cN_a\,\,\int_{\bR^{3,1}} (N_{aI} \tilde \Delta^I - M_a^I  \Delta_I) \wedge \text{Tr}(F_a).
\end{equation}
The axions and two-forms come in four-dimensional
Hodge dual pairs $(d\Upsilon^I, -d\Delta_I)$ and $(d\tilde\Upsilon_I,d\tilde \Delta^I)$, which can be derive from the ten-dimensional
Hodge duality $dC_3=-\star_{10} dC_5$.
One can show that the axions transform as 
\begin{equation}
\label{eqn:axion transformation}
\Upsilon^I \mapsto \Upsilon^I + \cN_a\, M_a^I\Lambda_a \qquad \text{and} \qquad \tilde \Upsilon_I \mapsto \tilde\Upsilon_I + \cN_a\, N_{aI}\Lambda_a
\end{equation}
under $U(1)_a$, which clearly leaves the four-dimensional couplings \eqref{eqn:4d terms} invariant. It was shown in \cite{Blumenhagen:2002wn}
that this gauging of the axionic shift symmetry precisely cancels the abelian and mixed anomalies.

Another important fact is that the couplings of the form $\int \Delta \wedge \text{Tr}(F_a)$ and $\int \tilde \Delta \wedge \text{Tr}(F_a)$
generically give rise to a Stuckelberg mass term for the $U(1)_a$ gauge bosons. Since we are dealing with orientifold
compactifications, there are also image branes on $\pi_a'$ with field strength $-F_a$, so that the relevant
couplings take the form
\begin{equation}
\label{eqn:BwedgeF}
\cN_a\,\,\int_{\bR^{3,1}} (N_{aI} - N_{aI}') \tilde \Delta^I \wedge \text{Tr}(F_a)\,\,\,-\,\,\, \cN_a\,\,\int_{\bR^{3,1}} (M_a^I - {M'}_a^I)  \Delta_I \wedge \text{Tr}(F_a).
\end{equation}
Though the $U(1)_a$ gauge bosons receive a mass, no symmetries of the action are broken, and so the gauge symmetry
selection rules associated with the $U(1)_a$ gauge symmetry survive in the low energy effective action as global selection
rules. These are precisely the global $U(1)$ symmetries which forbid superpotential terms in string perturbation theory.

However, as is fortunate for phenomenological purposes, it is often the case that some linear combination of the $U(1)_a$
gauge symmetries remains massless. As a condition on homology, this means that a linear combination $\sum_x \,q_x\, U(1)_x$
is massless when
\begin{equation}
\sum_x \cN_x\,q_x\,\,(\pi_x-\pi_x')=0.
\end{equation}	
As with the case of the condition on homology for tadpole cancellation, this can be intersected with a cycle $\pi_a$ wrapped
by a D6-brane to give constraints on the allowed forms of chiral matter. Using Table \ref{table:spectrum}, these constraints
are given by
\begin{equation}
\label{eqn:chiral masslessness constraint}
-q_a\cN_a\,\,(\#(\Ysymm_a) + \#(\Yasymm_a)) + \sum_{x\ne a} q_x \cN_x \,\, (\#(a,\ov x) - \#(a,x)) = 0,
\end{equation}
which becomes
\begin{equation}
\label{eqn:chiral masslessness constraint N1}
-q_a \,\,\frac{\#(a) - \#(\ov a) + 8 \#(\Ysymm_a)}{3} + \sum_{x\ne a} q_x \cN_x \,\, (\#(a,\ov x) - \#(a,x)) = 0,
\end{equation}
for the special case $\cN_a=1$. Any linear combination of $U(1)$'s which satisfies these conditions is an anomaly-free
$U(1)$ with a massless gauge boson. When trying to realize the standard model in D-braneworlds, there must
be such a $U(1)$ which allows for an interpretation as hypercharge. The particular linear combination corresponding
to hypercharge, sometimes called a ``hypercharge embedding", has important implications for the realization of MSSM
matter fields, and thus also for the structure of couplings.

\section{Concrete Model-Building Example: Toroidal Orbifolds \label{sec:toroidal orbifold}}
The beautiful geometric picture of particle physics offered by type IIa intersecting braneworlds has
encouraged much work in model building. Many of these examples are compactified on various
toroidal orbifolds\footnote{For recent work on global toroidal orbifold models and other recent work, see \cite{Forste:2010gw} and references therein. For systematic work on the landscape of string vacua for particular toroidal orbifolds, see \cite{Blumenhagen:2004xx}.}, which offer two distinct advantages over more generic Calabi-Yau backgrounds. In
particular, the homology cycles on a toroidal orbifold are particularly easy to visualize, which makes
model building a bit more intuitive. Furthermore, toroidal orbifolds offer a CFT description, and therefore
all of the power of vertex operator formalism can be brought to bear.

Generically, the toroidal orbifold is a six-torus modded out by a discrete group $\Gamma$, so that $\cM = T^6/\Gamma$.
We think of a factorizable six-torus $T^6=T^2\times T^2\times T^2$. One might wonder
about the simplest possibility, where $\Gamma$ is trivial and therefore we simply have the type IIa string
compactified on an orientifold of $T^6$. Unfortunately, due to simple considerations from the supersymmetry
condition, these models cannot realize the MSSM. In the literature, therefore, $\Gamma$ is non-trivial and
is usually of the form $\bZ_N$ or $\bZ_N\times \bZ_M$.

Before discussing the effects of $\Gamma$-action on $T^6$, it is necessary to mention another detail or two about
the orientifold. Introducing complex coordinates $z^i=x^i + i\,y^i$ on each of the $T^2$'s, the anti-holomorphic
involution acts as $\ov \sigma:z^i\mapsto \ov z^i$. On each $T^2$ there are exactly two different choices for the complex
structure which are consistent with the involution. They correspond to the torus and the tilted torus. The twofold
basis of one-cycles for the torus and tilted torus are $([a^i],[b^i])$ and $([a'^i],[b^i])$, where $[a'^i]=[a^i]+\frac{1}{2}[b^i]$.
Since the six-torus is factorizable, the three-cycles can be written as a product of three one-cycles as
\begin{equation}
\label{eqn:one-cycle param}
	\pi_a = \prod_{i=1}^3 (n_a^i[a^i]+\tilde m_a^i [b^i]),
\end{equation}
where $\tilde m_a^i = m_a^i$ for untitled tori and $\tilde m_a^i = m_a^i + \frac{1}{2} n_a^i$ for tilted tori.
Using the fact that the only non-vanishing intersection of one-cycles is $[a^i]\circ [b^i] = -1$, it is straightforward
to calculate
\begin{equation}
I_{ab} = \prod_{i=1}^3(n_a^i \tilde m_b^i - \tilde m_a^i n_b^i) = \prod_{i=1}^3(n_a^i m_b^i - m_a^i n_b^i).
\end{equation}
One should make careful note that the intersection number \emph{does not} depend on the choice of tilted or untilted
tori. This makes sense, of course, because topological quantities such as $I_{ab}$ must not depend on metric-related issues,
such as complex structure moduli.

Since we have specified a manifold on which to compactify, it is useful to recast the generic tadpole cancellation conditions
\eqref{eqn:tadpole} in terms of the wrapping numbers $(n,m)$. Independent of the tilt on each $T^2$, the O6-plane is wrapping
the cycle $2[a^i]$, so that the entire three-cycle reads $\pi_{O6}=8[a^1][a^2][a^3]$. The action of
$\ov \sigma$ on a generic cycle is simply $(n^i,\tilde m^i) \mapsto (n^i,- \tilde m^i)$. Parameterizing
the cycles in terms of wrapping numbers as in \eqref{eqn:one-cycle param}, the RR tadpole cancellation conditions become
\begin{align}
\label{eqn:T6 orientifold tadpole}
[a^1][a^2][a^3]&: \qquad \sum_{a=1}^K N_a \prod_i n_a^i = 16 \notag \\
[a^i][b^j][b^k]&: \qquad \sum_{a=1}^K N_a n_a^i \tilde m_a^j \tilde m_a^k = 0 \qquad \text{with} \qquad i\ne j\ne k
\end{align}
One might wonder why there are no equations for the three-cycle basis components of the form $[b][b][b]$ or $[a][a][b]$.
This is because $\tilde m^i\mapsto -\tilde m^i$ under $\ov \sigma$ kills any contribution to a component with an odd number of $b$'s.

As type IIa compactified on an orientifold of $T^6$ cannot realize the MSSM, it is important to examine the possibility
of non-trivial $\Gamma.$ Here we consider a well-studied choice for the orbifold group, where $\Gamma=\bZ_2\times\bZ_2$.
The generators of the $\bZ_2\times\bZ_2$ are given by $\omega$ and $\theta$, defined to be
\begin{equation}
	\omega: (z^1,z^2,z^3) \mapsto (-z^1,-z^2,z^3) \qquad \qquad \theta: (z^1,z^2,z^3) \mapsto (z^1,-z^2,-z^3). 
\end{equation}
Since $\Gamma$ also acts on the homology cycles of $T^6$, the simplification of the tadpole conditions \eqref{eqn:tadpole}
in terms of wrapping numbers must take this into account. In fact, this can be done for any $\Gamma$. We refer the interested
readers to the reviews \cite{Blumenhagen:2005mu,Marchesano:2007de,Blumenhagen:2006ci} for the derivation and expressions of the orbifold tadpole conditions.

\TABLE{
	\label{table:pati-salam example}
	\begin{tabular}{|c|c|c|}
		\hline Stack & N & $(n^1,m^1) \times (n^2,m^2) \times (n^3, m^3)$  \\ \hline \hline
		a & 8 & $(0,-1) \times (1,1) \times (1,1)$\\
		b & 4 & $(3,1) \times (1,0) \times (1,-1)$\\
		c & 4 & $(3,-1) \times (0,1) \times (1,-1)$\\
		1 & 2 & $(1,0) \times (1,0) \times (2,0)$\\
		2 & 2 & $(1,0) \times (0,-1) \times (0,2)$\\
		3 & 2 & $(0,-1) \times (1,0) \times (0,2)$\\
		4 & 2 & $(0,-1) \times (0,1) \times (2,0)$ \\\hline
	\end{tabular}
	\caption{Toroidal orbifold wrapping numbers for a Pati-Salam model with $U(4)_a\times U(2)_b \times U(2)_c \times U(1)_1 \times U(1)_2 \times U(1)_3 \times U(1)_4$ gauge symmetry. $N$ is twice what one might expect, due to the orbifold action.}
}

As an example, we present the wrapping numbers for a globally consistent model of \cite{Cvetic:2004ui} in table \ref{table:pati-salam example}
. This model is a type
IIa orientifold on $T^6/(\bZ_2\times\bZ_2)$ with intersecting D6-branes and Pati-Salam $SU(4)_C\times SU(2)_L\times SU(2)_R$
gauge symmetry after the Green-Schwarz mechanism has given masses to $U(1)$ gauge bosons. From the point of view of bifundamental matter under the Pati-Salam group, the chiral spectrum is particularly
nice, as it contains three families of $(4,\ov 2, 1)$ and $(\ov 4, 1,2)$. However, in addition to the Pati-Salam gauge symmetry, 
which arises from stacks $a$, $b$, and $c$, ``filler" branes labeled with integers are needed to satisfy the tadpole conditions.
This gives rise to many chiral exotics arising at the intersection of a filler brane with a Pati-Salam brane. The appearance
of chiral exotics at intersections with filler branes occurs
somewhat frequently in type II orientifolds, and often spoils the phenomenology.

\section{Perturbative Yukawa Couplings \label{sec:couplings}}
To this point, we have reviewed how low energy effective theories with particular gauge symmetry and chiral matter
arise in the context of type II orientifold compactifications. While these effects are the most important if a string
vacuum is to realize the particle physics of our world, it is also crucial that the couplings of the low energy theory
reproduce the structure in the Standard Model. This is a particularly important detail to investigate in the context of
intersecting brane models, as the gauge symmetries whose gauge bosons are given a Stuckelberg mass via the Green-Schwarz
mechanism impose global selection rules on couplings, forbidding crucial superpotential terms in string perturbation theory.
If a model is to realize such a forbidden, but desired, coupling, it must be due to a non-perturbative effect.

\subsection{Yukawa Couplings from String Amplitudes}

Before we address non-perturbative effects, we address those Yukawa couplings which \emph{are} present in string perturbation
theory, and thus give the leading order effects. To determine the structure of a given Yukawa coupling, one must calculate the
relevant correlation function in conformal field theory. We take the example of an up-flavor quark Yukawa coupling, which appears
in the superpotential as $H_u\,Q_L\, u_R$. It is phenomenologically preferred that at least one such Yukawa coupling be present
in string perturbation theory, as the top-quark Yukawa coupling is $\cO(1)$, which is difficult to obtain via a non-perturbative
effect. In terms of the vertex operators presented in section \ref{sec:CFT}, the relevant correlator is
\begin{equation}
\label{eqn:upyukawa}
\langle V_{-1}^{H_u} \,\,\, V_{-1/2}^{Q_L} \,\,\, V_{-1/2}^{u_R}\rangle,
\end{equation}
where the $NS$ and $R$ sector vertex operators are chosen so that we are using the bosonic component of the Higgs supermultiplet,
and the fermionic components of the quarks.
\FIGURE{
 \centering
 \includegraphics[scale=.5]{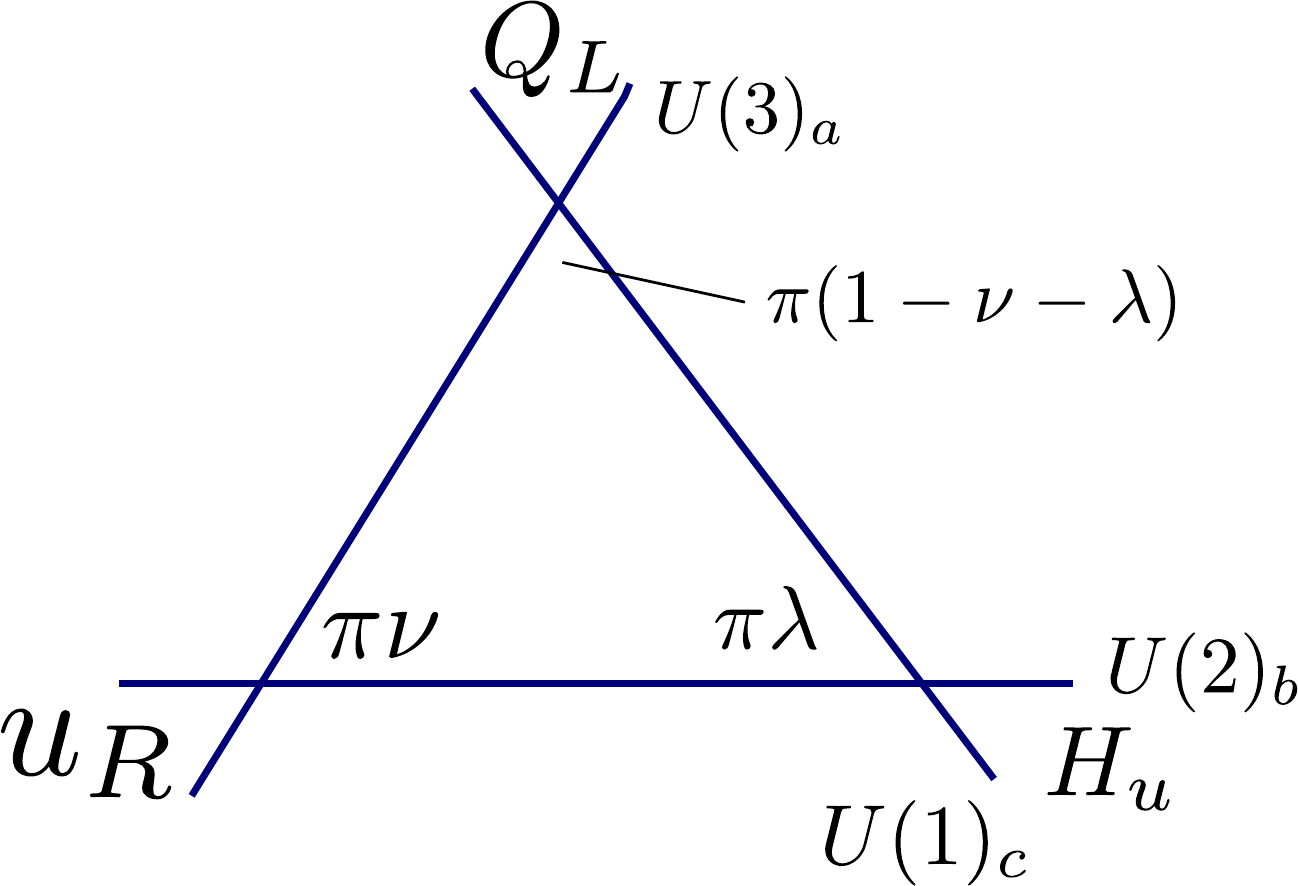}
 \caption{\small A two-dimensional slice of the Calabi-Yau graphically depicting, an up-flavor quark Yukawa coupling. The area of the triangle (worldsheet instanton) affects the scale of the coupling.}
 \label{fig:upyukawa}
}

Before being concerned about the precise structure of the correlator, the most coarse thing that one can do
is determine whether or not the operator is forbidden by symmetries in string perturbation theory. This depends
entirely on how the $H_u$, $Q_L$, and $u_R$ fields are represented in the brane stacks. For example, consider the
case of a three stacks of D-branes with $U(3)_a\times U(2)_b\times U(1)_c$ gauge symmetry, and that the linear combination
$U(1)_Y = \frac{1}{6}U(1)_a + \frac{1}{2}U(1)_c$ is left massless by the Green-Schwarz mechanism, which
we identify with hypercharge. Now suppose that the fields are realized as
\begin{equation}
H_u\sim (b,c)\qquad Q_L^1\sim (a,b)\qquad Q_L^2\sim (a, \ov b)\qquad u_R\sim (\ov a, \ov c),
\end{equation}
where two families of the left-handed quark doublets are realized as $Q_L^1$ and one as $Q_L^2$. Then the possible
up-flavor quark Yukawa couplings have $U(1)$ structure
\begin{equation}
	H_u\,Q_L^1\,u_R:\,\,\,(0,2,0) \qquad\text{and}\qquad H_u\,Q_L^2\,u_R:\,\,\,(0,0,0), 
\end{equation} 
the first of which has non-zero global $U(1)$ charge and is therefore forbidden in string
perturbation theory. In this case we have one family of up-flavor quarks perturbatively allowed and
two disallowed, which gives a nice explanation of the large top-quark mass. However, if the
model is to be phenomenologically viable, non-perturbative effects \emph{must} generate Yukawa couplings
for the other two families, otherwise the up-quark and charm-quark will be massless.

Since the vertex operators for each of the relevant fields is known explicitly, the correlator
\eqref{eqn:upyukawa} can be calculated explicitly using CFT techniques \cite{DFMS,Cvetic:2003ch,Abel:2003vv,Bertolini:2005qh}.
The non-trivial aspect of the calculation of this correlation function
involves calculating the three-point correlator of the bosonic twist fields which take 
into account the boundary conditions associated with the angles between branes. Having the
picture of a toroidal orbifold in our head, the target space picture of this Yukawa coupling
for one of the tori looks like Figure \ref{fig:upyukawa},
where the corresponding bosonic twist field amplitude that needs to be calculated is
\begin{equation}
\label{eqn:three twist}
\langle \sigma_\nu(z_1)\,\sigma_{-\nu-\lambda}(z_3)\,\sigma_\lambda(z_4)\rangle.
\end{equation}
Calculation of this correlator can be performed by calculating the four point disk correlator
\begin{equation}
\label{eqn:four twist}
\langle \sigma_{\nu}(z_1)\,\sigma_{-\nu}(z_2)\,\sigma_{-\lambda}(z_3)\,\sigma_{\lambda}(z_4)\rangle
\end{equation}
and extracting \eqref{eqn:three twist} in the limit where $z_2\mapsto z_3$. The spacetime picture of the
four point correlator is the blue trapezoid in Figure \ref{fig:fourpoint}, and the geometric picture of
the $z_2\mapsto z_3$ limit is to take the uppermost brane north, past the point of convergence of
the dotted red lines. For technical details on the calculation of these correlators, 
we refer the reader to \cite{Cvetic:2003ch}.

The twist field correlator determines the angular dependence of the Yukawa coupling, which is
often referred to as the ``quantum" part of the coupling, due to its dependence on CFT quantum correlators.
The explicit structure of a Yukawa coupling in a $T^6=T^2\times T^2\times T^2$ background is given by
\begin{equation}
h = \sqrt{2} \, g_s \,  2\pi \,  \prod_{I=1}^3 (\frac{16\pi^2\,\Gamma(1-\nu^I) \Gamma(1-\lambda^I)\,\Gamma(1-\nu^I-\lambda^I)}{\Gamma(\nu^I)\Gamma(\lambda^I)\Gamma(\nu^I+\lambda^I)})^{\frac{1}{4}} \,\,\, \sum_m \text{exp}(\frac{-A_I^m}{2\pi\alpha '}),
\end{equation}
where the indices $I\in\{1,2,3\}$ are one for each two-torus and $A_I^m$ is the area of the $m$-th triangle (worldsheet
instanton \cite{Cremades:2004wa,Aldazabal:2000cn}) on the $j$-th two-torus. The factor involving the worldsheet instantons is often called the classical factor.

\subsection{Coupling Issues, Exemplified}

In our simple example in the previous section, we saw that two of the families of up-flavor quark Yukawa couplings were forbidden
in string perturbation theory, as $H_uQ_L^1u_R$ had non-zero global $U(1)$ charge. In certain scenarios
non-perturbative effects can generate the missing couplings, but in this case the issue could be avoided entirely if all three families
of left-handed quark doublets appear as $Q_L^2$ instead of $Q_L^1$, as in that case all of the up-flavor quark Yukawa couplings would
be perturbatively allowed. This depends heavily on how the chiral matter in a given orientifold
compactification is realized at the intersection of two branes, as that determines the structure
of the global $U(1)$ charges.
\FIGURE{
 \centering
 \includegraphics[scale=.5]{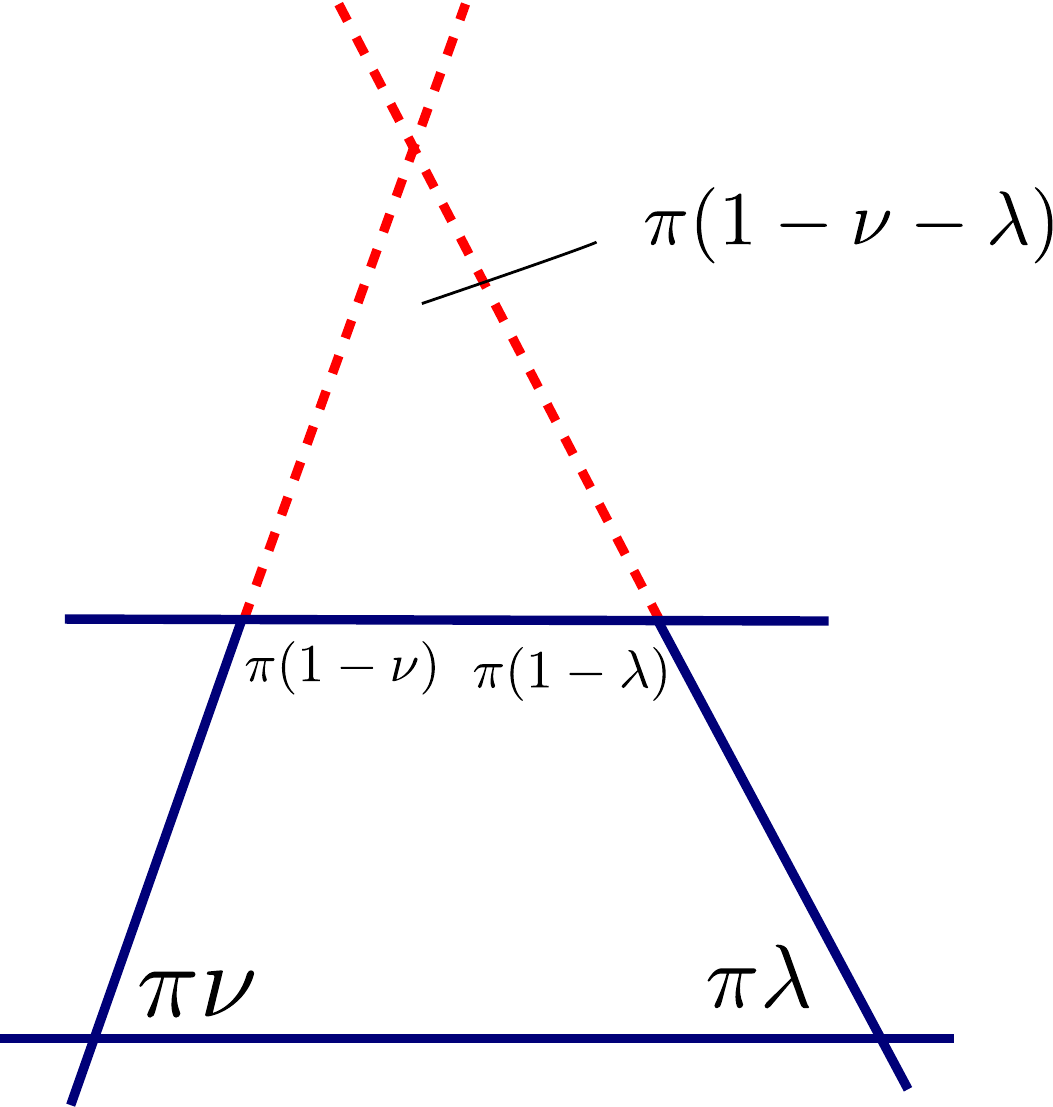}
 \caption{\small A spacetime picture of a four-point coupling. The three-twist correlator, relevant for Yukawa couplings, is extracted from the four-twist correlator appearing in this coupling.}
 \label{fig:fourpoint}
}
There are important phenomenological couplings which are $\emph{always}$ forbidden in string perturbation theory, though,
so that if a weakly coupled type II orientifold compactification is to realize them, it \emph{must} be at the non-perturbative level. In
this section, we discuss the non-perturbative generation of the always forbidden 
Majorana neutrino mass term  and its role in the seesaw mechanism. In addition, we discuss the non-perturbative generation of the always
forbidden $10\,10\,5_H$ Yukawa coupling, which gives mass to the top-quarks in Georgi-Glashow GUTs.

\vspace{.5cm}
\noindent \textbf{Example One: The Neutrino Masses} \\
\indent Consider a single Dirac neutrino mass coupling $h_\nu \,\, H_u \, L\, N_R$, which can be calculated as a function of moduli
using the conformal field theory techniques above. Then, after electroweak symmetry breaking, the Dirac mass term is 
$m_{D_\nu} = h_\nu \,\langle H_u \rangle$. Comparing to a generic quark mass $m_q = h_Q \, \langle H_u \rangle$, we see that
the quark masses and the neutrino masses are generically of the same order, unless the Yukawa couplings are tuned such that
$h_\nu \ll h_Q$.

While worldsheet instantons are able to account for the standard model fermion hierarchies to some degree, it is only in very
small regions of moduli space where they could account for the hierarchy $m_Q \sim 1\, GeV$ and $m_\nu\sim \, 10^{-3} \, eV$. As
in the particle theory literature, we would prefer to have some mechanism to account for this hierarchy, rather than attributing it
to some miraculous result of moduli stabilization.

One popular field theoretic mechanism which accounts for the small neutrino masses is the type I seesaw mechanism. In this mechanism,
in addition to the Dirac type neutrino mass term $h_\nu \, H_u\, L\, N_R$, there is a Majorana neutrino mass term of the form $M_R N_R N_R$. The 
neutrino mass matrix takes the form
\[ \left( \begin{tabular}{cc}
0 & $h_\nu \,\,\, \langle H_u \rangle$ \\
$h_\nu \,\,\, \langle H_u \rangle$ & $M_R$
\end{tabular} \right)\] 
giving rise to mass eigenvalues $M_R$ and $\frac{h_\nu^2 \langle H_u \rangle^2}{M_R}$ in the limit of large $M_R$. Thus, one of the
mass eigenvalues has been ``seesawed" to a very small value by the large Majorana mass, giving (in this mechanism) the reason for
the very small neutrino masses observed in nature.

While this mechanism is nice from the point of view of field theory, there is an important difficulty which arises when attempting
to realize a Majorana mass term in string theory. As the right-handed neutrinos $N_R$ will have global $U(1)$ quantum numbers in a type II orientifold compactification, the Majorana
mass term $M_R N_R N_R$ will also be charged with respect to the global $U(1)$'s, and is therefore forbidden in string perturbation theory.
Therefore, in type II orientifold compactifications it is difficult to account for the smallness of the neutrino masses in string perturbation theory: in the absence of extreme fine-tuning of the moduli, $m_{D_\nu}$ and $m_Q$ are of the same order, and the
seesaw mechanism cannot be realized, as the Majorana mass term is forbidden.

\vspace{.5cm}
\noindent \textbf{Example Two: the $10\, 10\, 5_H$ in $SU(5)$ GUTs} \\
\indent There is another well-known coupling problem that arises when trying to realize Georgi-Glashow
GUTs in weakly coupled type II orientifold compactifications. In these models, the $SU(5)$ gauge theory is realized by a stack of five
spacetime-filling D-branes wrapping a non-trivial cycle in the Calabi-Yau, and chiral matter charged under the $SU(5)$ factor is realized
at the intersection of this five-stack with some other D-brane. It is the possibility of symmetric and antisymmetric matter representations, in addition to the bifundamental, which allows for the realization of the standard $SU(5)$ GUT particle representations. Specifically, the $10$ representation of $SU(5)$
can be realized as $\Yasymm_5$ at the intersection of the five-stack with its orientifold image.

Though the proper spectrum can be realized, there is an immediate problem at the level of couplings. Since the $10$ is realized as an antisymmetric,
it comes with charge $2$ under the $U(1)$ of the five-stack. In addition, the $5_H$ comes with charge $1$, since it a fundamental of $SU(5)$, and
the top-quark Yukawa coupling $10\, 10\, 5_H$ has charge $5$ and is therefore \emph{always} forbidden in string perturbation theory. On the other hand, the 
bottom-quark Yukawa coupling $10 \, \ov 5 \, {\ov 5}_H$ can be present in string perturbation theory, giving a massive bottom-quark and a massless top-quark.
This inverts the standard hierarchy and is a major phenomenological pitfall that must be remedied if one hopes to realize realistic Georgi-Glashow
GUTs in weakly coupled type II orientifold compactifications.

\section{Non-Perturbative Superpotential Corrections: D-instantons \label{sec:instantons}}
We have now seen that it would be phenomenologically useful if some non-perturbative effect were able
to generate superpotential couplings which are forbidden in string perturbation theory. Doing so would
require that the non-perturbative physics somehow cancels the excess $U(1)$ charge associated
with a perturbative Yukawa coupling. A the reader familiar with the KKLT \cite{Kachru:2003aw} scenario (for example)
might be concerned that D-instantons cannot serve this purpose, as there a euclidean D3-instanton
in type IIb generated a non-perturbative correction without charged matter which was responsible for stabilizing
a K\" ahler modulus.
However, it was shown in \cite{Blumenhagen:2006xt,Ibanez:2006da,Florea:2006si} that in the presence of spacetime
filling gauge D-branes the axionic shift symmetries which are gauged by the Green-Schwarz mechanism
cause the axions to be charged with respect to the global $U(1)$ symmetries. These axions appear in instanton corrections and can cancel the net $U(1)$ charge of perturbatively forbidden couplings,
giving rise to these couplings at the non-perturbative level.

\FIGURE{
 \centering
 \includegraphics[scale=.4]{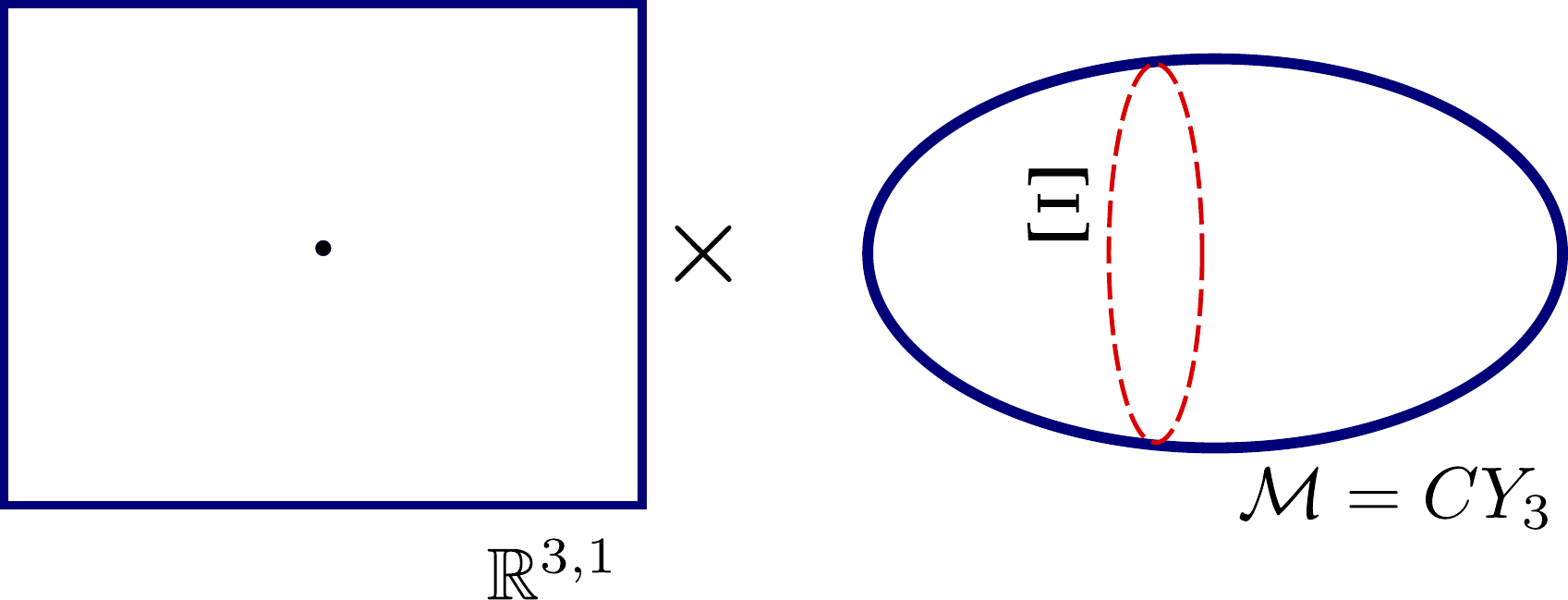}
 \caption{\small}
 \label{fig:instantion cycle}
 \caption{A target space view of a euclidean D-brane instanton, which is pointlike in spacetime and wraps a non-trivial
 cycle $\Xi$ in the Calabi-Yau.}
}

Consider a type II orientifold compactification with euclidean D-instantons in the background. The instantons
are pointlike in spacetime and wrap a non-trivial cycle in the Calabi-Yau. The instanton
action is of the form
\begin{equation}
S_{inst} = S^E_{cl} + S(\cM,\Phi)
\end{equation}
where $\cM$ are the set of instanton zero modes and $\Phi$ are the set of charged matter fields present in the 
low energy theory. The instanton correction to the low energy effective theory in four dimensions takes the form
\begin{equation}
\label{eqn:path integral}
S^{4d}_{np}(\Phi) = \int [\cD \cM ] \,\,e^{-S_{inst}},
\end{equation}
where the structure of the correction is determined by $\cM$ and $\Phi$, and its magnitude is set by the classical
instanton suppression factor, which depends on the volume wrapped by the instanton in the Calabi-Yau.

\subsection{Instanton Heuristics}

For the sake of concreteness, we again work in type IIa, where the instantons are euclidean D2 branes which are point-like in spacetime
and wrap non-trivial three-cycles in the Calabi-Yau. Taking a three-cycle $\Xi$, the classical action of the instanton is
\begin{equation}
S_{cl}^E = T_{E2}[\frac{1}{g_s}\int_\Xi \sqrt{\text{det}\, G} - i \int_\Xi C_3],
\end{equation}
where the first term comes from the Born-Infeld action and the second from the Wess-Zumino action. 
Here $e^{-Re\, S_{cl}^E}$ is a real suppression factor that sets the scale of superpotential corrections. Its
value is set by
\begin{equation}
Re \, S_{cl}^E = \frac{T_{E2}}{g_s} \, V_\Xi = \frac{8\pi^2}{g_a^2} \, \frac{V_\Xi}{V_{\pi_a}},
\end{equation}
and can lead to very large suppression due to its exponential nature. This is phenomenologically very
relevant, as (for example) it allows one to realize neutrino masses of the correct order with only a highly
suppressed Dirac term $LH_uN_R$ \cite{Cvetic:2008hi}, without resorting to the seesaw mechanism. Another alternative
to the seesaw mechanism includes the generation of the Weinberg
operator $LH_uLH_u$ by a D-instanton \cite{Cvetic:2010mm}.

One might recall from section \ref{sec:Generalized Green-Schwarz Mechanism} that dimensional reduction of
the Ramond-Ramond three-form $C_3$ on three cycles gives rise to four-dimensional axions $\Upsilon^I$ and
$\tilde \Upsilon_I$, which transform under $U(1)_a$ due to the generalized Green-Schwarz mechanism. These
axions enter the classical instanton action as
\begin{equation}
Im\, S_{cl}^E = T_{E2} \, \int_\Xi C_3 = T_{E2} \, (N_{\Xi I}\Upsilon^I - M_\Xi^I \tilde \Upsilon_I).
\end{equation}
The gauge field one-form associated with $U(1)_a$ transforms as $A \mapsto A + d\Lambda_a$, and the axions transform as
\begin{equation}
\Upsilon^I \mapsto \Upsilon^I + \cN_a \,\, M_a^I \Lambda_a \qquad \text{and} \qquad \Upsilon_I \mapsto \Upsilon_I + \cN_a \,\, N_{a,I} \Lambda_a
\end{equation}
from which it can be seen that the classical instanton action transforms as
\begin{equation}
e^{-S_{cl}} \mapsto e^{-S_{cl} + i \, Q_\Xi^a\,\, \Lambda_a},
\end{equation}
with $Q_\Xi^a=\cN_a\,\, \Xi\, \circ\, \pi_a$. 

Taking the orientifold and image branes into account induces extra shifts
in the classical instanton action, so that in full generality
\begin{equation}
\label{eqn:instanton charge}
Q_\Xi^a=\cN_a\,\Xi\circ(\pi_a - \pi_a ').
\end{equation}
This is precisely the net charge carried microscopically by the charged instanton zero modes in the path integral measure, as will be discussed in section \ref{sec:fermionic zero modes}.
If a subset of matter fields $\phi_i\in\Phi$ has charges $Q_i$ such that $Q^a_\Xi + \sum_i Q_i^a = 0 \,\,\, \forall a$,
an instanton wrapped on $\Xi$ can, in principle, generate a superpotential coupling of the form
\begin{equation}
\label{eqn:correction}
e^{-S_{cl}^{E2}}\prod_i \phi_i,
\end{equation}
since the non-trivial transformation of the axions cancels the global $U(1)$ charge of $\prod_i \phi_i$.

\subsection{Fermionic Zero Modes \label{sec:fermionic zero modes}}
The arguments of the previous section heuristically showed that the gauging of shift symmetries by the Green-Schwarz mechanism make it possible for couplings of the form \eqref{eqn:correction} to be gauge invariant, depending on the $U(1)$ charges of the matter fields. Whether or not such a a term is actually generated by an instanton depends heavily on the microscopic properties of the instanton, in particular its fermionic zero modes, which correspond to massless open strings.

The importance of fermionic zero modes for determining non-perturbative corrections is well known in other areas of the landscape. For example, in \cite{Witten:1996bn} Witten argued that an M5-instanton wrapped on a $6$-cycle
$\Xi_{M5}$ must satisfy
\begin{equation}
	\chi(\Xi_{M5},\cO_{\Xi_{M5}}) = \sum_{i=0}^3 h^i(\Xi_{M5},\cO_{\Xi_{M5}}) = 1
\end{equation}
if it is to contribute to the superpotential. This is a constraint on particular (uncharged) fermionic zero modes
of the instanton, which are counted by the Hodge numbers $h^i(\Xi_{M5},\cO_{\Xi_{M5}})$. Similar constraints 
exist for the uncharged modes in type II, as, for example, a euclidean D3-instanton wrapped on a holomorphic
divisor $D$ must satisfy $\chi(D,\cO_D)=1$ if it is to contribute to the superpotential.

The influence of the fermionic modes on non-perturbative corrections is easy to see: the path integral
integral is over all fermionic zero modes, so if these modes are not lifted or the instanton action does
not have appropriate terms for soaking them up, then the Grassman integral evaluates to zero. Suppose
(in an unrealistic but illustrative example) that $S(\cM,\Phi)= a\,\xi + b\,\eta + c\,\xi\eta$, where
the Greek variables are fermionic zero modes. Then from \eqref{eqn:path integral} the non-perturbative
correction would be
\begin{equation}
	\label{eqn:fermionic modes example}
	e^{-S_{cl}^{E}} \int [d\xi] [d\eta] \,\,\, e^{-(a\,\xi + b\,\eta + c\,\xi\eta)} = 
	e^{-S_{cl}^{E}} \int [d\xi] [d\eta] \,\,\, (1-(a\,\xi + b\,\eta + c\,\xi\eta)) =
	c \,\, e^{-S_{cl}^{E}}.
\end{equation}
It is easy to see that if the third term in $S(\cM,\Phi)$ were not present, then the correction would
be absent.

One might intuitively think that, since the basic branes in our theory are gauge D-branes and euclidean
D-instantons, there exist two types of fermionic zero modes living in the instanton worldvolume, corresponding
to a string from the instanton to itself and a string from the instanton to a gauge brane. Indeed, this is
the case, as can be shown (for example) by CFT techniques. The strings from the instanton to itself are
known as uncharged zero modes, and are the modes counted by $h^i(D,\cO_D)$ in type IIb\footnote{ For a beautiful explanation of how these modes lift to modes of vertical M5 instantons in F-theory, see \cite{Blumenhagen:2010ja}, and for a lift to F-theory of an instanton generating the $10\, 10\, 5_H$ see \cite{Cvetic:2010rq}. For a generic review of ED3 zero modes from the point of view of sheaf cohomology, see section 2 of \cite{Cvetic:2010ky}.} which contribute
to the holomorphic genus. The strings from the instanton to a gauge brane are charged under the gauge group
of the D-brane, and thus are known as charged modes.

\subsubsection{Uncharged Zero Modes}

Perhaps the most crucial of the uncharged modes are the ones associated with the breakdown of supersymmetry.
Recall that type II string theory compactified on a Calabi-Yau manifold gives rise to $\cN=2$ supersymmetry
in four dimensions, an $\cN=1$ subalgebra of which is preserved by the orientifold, with supercharges
$Q^\alpha$, $\ov{Q}^{\dot \alpha}$. A spacetime-filling D-brane wrapping a 1/2-BPS cycle\footnote{These are special Lagrangians for D6-branes,
holomorphic divisors for D7-branes.}  might preserve the same supercharges $Q^\alpha, \ov Q^{\dot \alpha}$, in which
case the D-brane is supersymmetric with respect to the orientifold. The orthogonal complement to the $\cN=1$ algebra
preserved by the orientifold, which has supercharges $Q'^\alpha$ and $\ov Q'^{\dot \alpha}$,
then gives four Goldstinos associated with the four broken supersymmetries.

The key point is that, due to localization in the four extended dimensions, a 1/2-BPS D-instanton does not preserve the
four supercharges $Q^\alpha$, $\ov Q^{\dot \alpha}$ preserved by a gauge D-brane and the orientifold, 
but instead the combination $Q'^{\alpha}$ and $\ov Q^{\dot \alpha}$.
There are then four Goldstinos in the instanton worldvolume associated with the breakdown of supersymmetry: two chiral modes
$\theta^\alpha$ associated with the breaking of $Q^\alpha$, and two anti-chiral modes $\ov \tau^{\dot \alpha}$ \cite{Argurio:2007qk,Argurio:2007vqa} associated
with the breaking of $\ov Q'^{\dot \alpha}$. With the $\theta$ mode identified as the $\theta$ mode for $\cN=1$ theories in
four dimensions, the instanton might contribute a superpotential correction if the $\ov \tau$ modes are
somehow saturated or lifted.

There are numerous ways for this to happen. One common possibility is that the instanton wraps an orientifold invariant
cycle, in which case the $\ov \tau$ mode is projected out. In the case of a type IIa orientifold compactification, this can
be seen directly in the CFT formalism, where the vertex operators associated with the $\theta$ and the $\ov \tau$ mode are given by
\begin{align}
V^{\theta}_{-\frac{1}{2}} = \theta_\alpha \, e^{-\frac{\phi}{2}} \, S^\alpha(z) \,\Sigma_{\frac{3}{8},\frac{3}{2}}(z) \notag \\
V^{\ov \tau}_{-\frac{1}{2}} = {\ov \tau}^{\dot \alpha} \, e^{-\frac{\phi}{2}} \, S_{\dot \alpha}(z) \,\Sigma_{\frac{3}{8},-\frac{3}{2}}(z),
\end{align}
and the $\Sigma$ fields are spin fields describing fermionic degrees of freedom on the internal space. The subscripts of $\Sigma$ give the conformal dimension and worldsheet $U(1)$ charge. The orientifold projection induces extra constraints on the structure of Chan-Paton factors, which the $\ov \tau$ modes do not satisfy and therefore they are projected out.

From the point of view of type IIb compactified on a generic Calabi-Yau with a euclidean D3 instanton wrapping a 
holomorphic divisor $D$, the presence or absence of $\theta$ and $\ov \tau$ is counted by
the Hodge number $h^{0,0}(D)\cong h^0(D,\cO_D)$. The presence of the holomorphic $\bZ_2$-action $\sigma$ associated with the orientifold
allows for a decomposition of ordinary sheaf cohomology into a sum of $\bZ_2$-equivariant sheaf cohomology as
\begin{equation}
	H^i(D,\cO_D) \cong H^i_+(D,\cO_D) \oplus H^i_-(D,\cO_D),
\end{equation}
reflecting the fact that each of the zero modes transform with a sign under $\sigma$. The $\ov \tau \in h^0_-(D,\cO_D)$ mode is the one which transforms with
a $-$ sign, and is therefore absent in the case of an orientifold invariant divisor. Such an instanton is called an $O(1)$ instanton.

These are not the only uncharged zero modes, however. For example, in addition to the $\theta$ and $\ov \tau$ Goldstino modes 
associated with the breakdown of supersymmetry due to the localization of the instanton in four-dimensional spacetime, 
there are additional uncharged modes corresponding to the localization of the instanton on submanifolds of the Calabi-Yau.
These deformation modes also admit both a CFT description, when available, and a description in terms of cohomology. The latter
can be seen in type IIb as deformations of a holomorphic divisor $D$, which are given by global sections of the normal bundle, and
thus are counted by $h^0(D,N_{D|\cM})$\footnote{Deformation modes, as well as other uncharged modes, are often said to be counted by
$h^i(D,\cO_D)$ in the literature, which may not be the most illuminating presentation. To help motivate this, note that in a 
Calabi-Yau manifold a holomorphic divisor has $K_D = N_{D|\cM}$,
so that by Serre duality we have $H^0(D,N_{D|\cM})\cong H^2(D,K_D\otimes N_{D|\cM}^*)\cong H^2(D,\cO_D)$. Therefore, by some
simple isomorphisms, the intuitive notion of deformation modes as normal bundle sections is recast as $\cO_D$ sheaf cohomology.}. If a cycle has no deformation modes, it (as well as an instanton wrapping it) is said to
be rigid. 

In order for an instanton to contribute to the superpotential, it must realize the $\theta$ mode and none of the other
uncharged modes\footnote{In type IIb, the precise statement in cohomology is $h^0_+(D,\cO_D)=1$ and all others zero. This necessary
and sufficient constraint automatically satisfies the necessary constraints $\chi(D,\cO_D)=1$, as it must.}, as the superpotential appears in the $\cN=1$ spacetime action as $\int d^4x\, d^2\theta \, W(\Phi)$. There are at least two possible reasons for the absence of a zero mode. First, it may be absent to begin
with due to being projected out. This is the reason for the absence of the $\ov \tau$ mode in the case of an $O(1)$ instanton.
The second possibility is that the extra zero modes are ``saturated" or ``soaked up". This is the case in \eqref{eqn:fermionic modes example},
where the $\chi$ and  $\eta$ integrals evaluate to one when integrating over the $c\,\chi\eta$ term.

\subsubsection{Charged Zero Modes}
\TABLE{
	\label{table:charged modes}
	\begin{tabular}{|c|c|c|}
		\hline Zero Mode & Representations & Number \\
		\hline $\lambda_a\equiv \lambda_{\cE,\cD}$ & $(-1_\cE,\fund_\cD)$ & $I_{\cE,\cD}^+$ \\
		$\ov\lambda_a\equiv \lambda_{\cD,\cE}$ & $(1_\cE,\antifund_\cD)$ & $I_{\cE,\cD}^-$ \\ \hline
	\end{tabular}
	\caption{Representations and multiplicity of charged modes appearing at the intersection of an instanton $\cE$ and a gauge
	brane $\cD$.}
}
The charged zero modes are massless strings from a euclidean D-instanton to a gauge D-brane, which are therefore charged
under the gauge group of the D-brane. They are the microscopic modes that carry the global $U(1)$ charges which compensate
for the overshoot in $U(1)$ charge of perturbatively forbidden couplings. The form of the superpotential corrections
involving charged matter depends heavily on these charged modes and is calculated using the instanton calculus presented
in \cite{Blumenhagen:2006xt} and reviewed in \cite{Blumenhagen:2009qh}. In short, it tells one how to determine the structure of $S(\cM,\Phi)$ in the
instanton action based on CFT disc diagrams involving charged matter fields and charged zero modes. For the sake of brevity, 
we refer the reader to those sources for a general discussion of instanton calculus, and instead we present the CFT basics of charged zero modes and an illustrative
example that makes the physics and basics of the methods very clear.

The Ramond sector open string vertex operator corresponding to a charged matter mode between a brane $\cD_a$ and an instanton $\cE$ is given by
\begin{equation}
V^{\lambda_a^i}_{-\frac{1}{2}}(z) = \lambda_a^i e^{-\frac{\phi}{2}}\Sigma_{\frac{3}{8},-\frac{1}{2}}^{D_a,\cE}(z)\sigma_{h=1/4}(z),
\end{equation}
where $i$ is the gauge index of the brane $\cD_a$ and the $\sigma$ fields are the 4D spin fields arising from the twisted 4D worldsheet bosons carrying half-integer modes. Due to the four Neumann-Dirichlet boundary conditions between the brane and the instanton in $\bR^{3,1}$, the zero point energy in the NS-sector is shifted by $L_0=1/2$. This makes all NS-sector states massive, so that the only charged zero modes come from the Ramond sector. We emphasize that the net $U(1)$ charge of these zero modes, as dictated by Table \ref{table:charged modes}, is precisely equivalent to the charge of the classical instanton action \eqref{eqn:instanton charge}.

We examine the up-flavor quark sector of the model presented in Table 1 of \cite{Cvetic:2009ez}\footnote{In fact this model is a quiver,
not a globally defined orientifold compactification. See section \ref{sec:quivers} for more information.}. That model is a four-stack
quiver with $U(3)_a\times U(2)_b \times U(1)_c \times U(1)_d$ gauge symmetry which becomes $SU(3)_C\times SU(2)_L \times U(1)_Y$
due to the Green-Schwarz mechanism, with the Madrid hypercharge embedding $U(1)_Y=\frac{1}{6}U(1)_a + \frac{1}{2}U(1)_c + \frac{1}{2}U(1)_d$.
The fields relevant for up-flavor quark Yukawa couplings are realized as
\begin{equation}
H_u: \,\,\, (b,c) \qquad Q_L: \,\,\, (a,\ov b) \qquad u_R^1: \,\,\, 1\times(\ov a, \ov c) \qquad u_R^2: \,\,\, 2\times(\ov a, \ov d),
\end{equation}
so that the Yukawa couplings have global U(1) charge
\begin{equation}
H_u\,Q_L\,u_R^1: (0,0,0,0) \qquad \qquad H_u\, Q_L \, u_R^2: (0,0,1,-1).
\end{equation}
Since $u_R^2$ has multiplicity two, two families are perturbatively forbidden, while one family is perturbatively realized,
as one might hope given the hierarchy of the top-quark mass relative to the up and the charm.
\FIGURE{
 \centering
 \includegraphics[scale=.7]{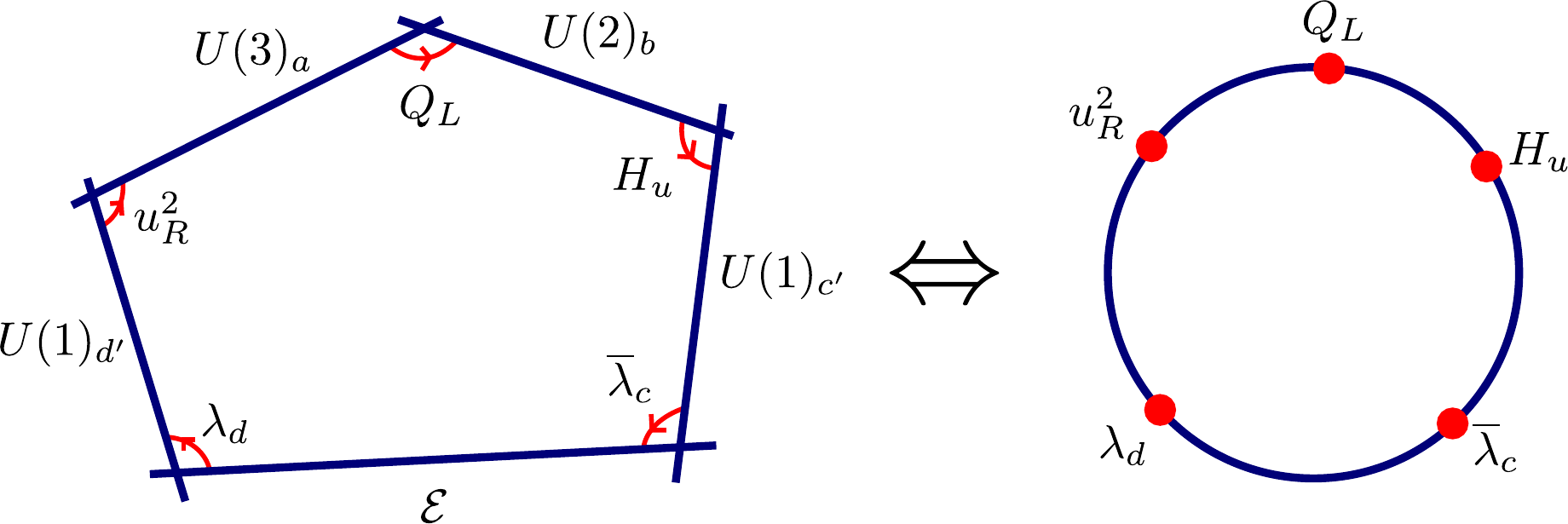}
 \caption{\small}
 \label{fig:spacetimeanddisk}
 \caption{A spacetime view of a non-perturbatively generated up-flavor quark Yukawa coupling, along with the corresponding disc
 diagram which contributes to the instanton action.}
}

To generate the missing $H_uQ_Lu_R^2$ couplings in type IIa, an instanton $\cE$ would have to exhibit intersection numbers
\begin{equation}
I_{\cE a}=0 \qquad I_{\cE b}=0 \qquad I_{\cE c} = 1 \qquad I_{\cE d} = -1,
\end{equation}
which gives rise to two charge modes, $\ov \lambda_c$ and $\lambda_d$. One can heuristically ``see" that this cancels the excess global
$U(1)$ charge in Figure \ref{fig:spacetimeanddisk} by the fact that spacetime picture is closed and the arrows point in in a consistent
direction. The corresponding disk diagram, also drawn in the figure, contributes the instanton action.

If, in a global embedding which realizes this
spectrum, a rigid O(1) instanton exists with this intersection pattern, then one can perform the instanton calculus with the mentioned
disk contribution.
The path integral \eqref{eqn:path integral} takes the form
\begin{align}
\int d^4x\,d^2\theta\,d\ov \lambda_c\, d\lambda_d \,\,\,e^{-S_{cl}^\cE +Y^J \, \ov \lambda_c \, H_uQ_Lu_R^{2,J} \, \lambda_d} &= \notag \\
e^{-S_{cl}^\cE} \,\,\, \int d^4x\,d^2\theta\,d\ov \lambda_c\, d\lambda_d \,\,\, Y^J \, \ov \lambda_c \, H_uQ_Lu_R^{2,J} \, \lambda_d &= \notag \\
e^{-S_{cl}^\cE} \,\,\, \int d^4x\,d^2\theta \,\,\, Y^J \, H_uQ_Lu_R^{2,J},
\end{align}
where $J$ runs across the two family indices for $u_R^2$.
In such a case the up-quark and charm-quark masses are suppressed by a factor of $e^{-S_{cl}^\cE}$ relative to the top-quark mass. In principle
$Y^J$ can give a hierarchy between the up-quark and charm-quark, since it depends on worldsheet instantons, but this generically depends
heavily on the details of moduli stabilization.

\section{Braneworld Quivers: The Bottom-Up Approach \label{sec:quivers}}
To this point we have discussed how all of the basic ingredients of real-world particle physics can be realized in the context
of weakly coupled type II orientifold compactifications. In particular, gauge symmetry lives on the worldvolume of
spacetime filling D-branes with possible gauge groups $U(N)$, $Sp(2N)$, or $SO(2N)$. Chiral matter appears at the intersection
of two stacks of D-branes, with the type and amount of chiral matter dictated by the topological intersection numbers
of the D-branes. Finally, the presence or absence of superpotential couplings depends crucially on the charge of couplings under
the $U(1)$ symmetries associated with the $U(N)$ branes.

D-braneworlds offer a beautiful
geometric picture which suggests the possibility of arranging branes in such a way that something very similar to the MSSM
is obtained. A top-down approach would first require specifying a Calabi-Yau manifold $\cM$ together with a $\bZ_2$ involution
on the space, which would allow for the identification of O-planes and follow with an investigation of the types of arrangements
of D-branes allowed by tadpole cancellation. Perhaps with the specification of further data (e.g. fluxes in type IIb, for chirality),
the massless spectrum can be calculated and the global $U(1)$ charges of the matter can be determined, allowing for the determination
of perturbative superpotential couplings. One can then perform a scan of possible instanton cycles to determine which might have
the proper fermionic zero mode structure for superpotential contribution. Needless to say, this quickly becomes rather involved.

Though a ``top-down" model is necessary if string theory is to provide the correct description of particle physics
in our world, this does not necessarily mean that the best way to identify promising models is by taking a top-down approach
to each vacuum. Recently, a ``bottom-up" approach \cite{Antoniadis:2000ena,Aldazabal:2000sa,Antoniadis:2001np} has emerged which suggests looking at certain subsets
of data associated with a string vacuum, with the hope that one can say non-trivial things across broader patches of the
landscape, despite the fact that certain details have been ignored. This approach is only good to the extent that
the ignored details don't destroy the physics determined by the subset of vacuum data of interest.

We already saw an example of this approach in section \ref{sec:CFT}, when looking at the up-quark Yukawa couplings $H_u\,Q_L^1\, u_R$
and $H_u\, Q_L^2\, u_R$, where the fields are realized by three-stacks of D-branes. Notice that we specified neither a Calabi-Yau
manifold nor a $\bZ_2$ involution in this example, and yet were able to make statements about couplings based on assumptions
about how chiral MSSM matter is realized at the intersection of various brane stacks. This information would be a subset of the
information associated with the quiver realized in a type II orientifold compactification.

In fact, much has been learned recently about particle physics in type II by studying quivers, which are a subset of the data defining
a type II orientifold compactification. Generically, a quiver is made of nodes and edges between them, where the nodes represent gauge D-branes and an
edge represents matter at the intersection of the corresponding D-branes. Thus, a quiver encodes the gauge symmetry and matter content in a type II orientifold compactification, including global $U(1)$ charges. We emphasize that these \emph{are not} globally consistent string compactifications. They cannot be, as quivers are only a subset of the data associated with a string vacuum. However, a given quiver can be shown to be \emph{compatible} with global consistency if it satisfies the necessary conditions \eqref{eqn:chiral tadpole constraint} and \eqref{eqn:chiral masslessness constraint}, which do contain genuinely string constraints that are not already present in field theory.

An example of a consistent type II quiver which realizes the exact
MSSM spectrum is given in Figure \ref{fig:quiver}, which corresponds to MSSM matter being realized as
\begin{align}
Q_L: \,\,\,2\times(a,\ov b), \,\,\,1\times(a,b) \qquad\,\,\, u_R:\,\,\,3\times(\ov a, \ov c) \qquad\,\,\, d_R:\,\,\, 3\times \Yasymm_a \notag \\ \notag \\ 
L: \,\,\,3\times(b,\ov c) \qquad\,\,\, E_R: \,\,\,3\times \Ysymm_c \qquad\,\,\, H_u: \,\,\,1\times (b,c) \qquad\,\,\, H_d:\,\,\, 1\times(\ov b, \ov c), \notag
\end{align}
where the gauge symmetry is $U(3)_a\times U(2)_b\times U(1)_c$ and the massless hypercharge is realized as $U(1)_Y = \frac{1}{6} U(1)_a + \frac{1}{2} U(1)_b$.
\FIGURE{
 \centering
 \includegraphics[scale=1.3]{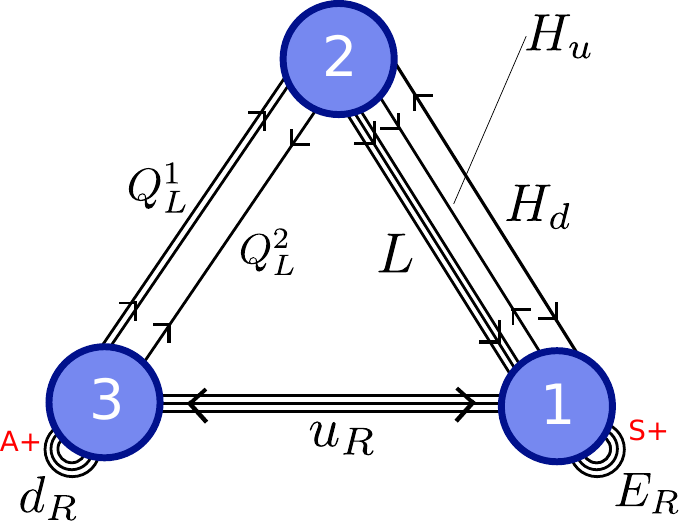}
 \caption{\small}
 \label{fig:quiver}
 \caption{A type II quiver which realizes the exact MSSM spectrum and satisfies the necessary constraints \eqref{eqn:chiral tadpole constraint}, \eqref{eqn:chiral masslessness constraint}, and \eqref{eqn:chiral masslessness constraint N1} for tadpole cancellation
 and a massless hypercharge.}
}
There are a few details required to understand the quiver diagram properly. First, since we are in the framework of type II orientifold compactifications,
there is an image brane associated with each $U(N)$ brane, which is all of the branes, in this case. Rather than doubling the number of nodes, we double the
number of arrows on the edges, with an arrow coming out of a node representing a fundamental, and an arrow going into a node representing an antifundamental.
Therefore, there are four options for arrow orientation on each edge, representing all possible bifundamental representations between those two branes and their
orientifold images. Second, edges from a node to itself implicitly correspond to a string stretching between a brane and its image, which yields a symmetric or
antisymmetric representation. In the $U(N)$ case, in addition to being symmetric or antisymmetric of $SU(N)$, there is a choice of $U(1)$ charge $\pm 2$, and so
we label these edges with either an $A$ or an $S$ and either a $+$ or a $-$.

This quiver contains enough information to say a great deal about the structure of the Yukawa couplings. Specifically, two of the families of up-quarks and
all of the leptons have perturbatively realized Yukawa couplings, while one of the up-quark families and all of the down-quark families are forbidden in
string perturbation theory. Note that the perturbatively forbidden up-flavor quark Yukawa coupling has global $U(1)$ charge $(0,2,0)$, as does the R-parity violating
coupling $LLE_R$. Therefore, any D-instanton which generates the missing up-flavor Yukawa coupling will also generate an R-parity violating operator at a very
high level. For this reason, and perhaps others, any type II orientifold compactification giving rise to this quiver is phenomenologically ruled out.

Any given quiver can, in principle, be realized in many different global embeddings, allowing one to make statements across broader patches of the landscape.
For example, in addition to the quiver presented in Figure \ref{fig:quiver}, there are 23 other three-stack quivers with $U(1)_Y = \frac{1}{6} U(1)_a + \frac{1}{2} U(1)_b$
and the exact MSSM spectrum that also satisfy the necessary constraints \eqref{eqn:chiral tadpole constraint} and \eqref{eqn:chiral masslessness constraint}. There is one other linear combination for the hypercharge which might realize the MSSM spectrum with three-stacks, which also gives
rise to a total of 24 quivers with the exact MSSM spectrum that satisfy \eqref{eqn:chiral tadpole constraint} and \eqref{eqn:chiral masslessness constraint}. One can therefore make the following statement: if the exact MSSM is to be realized in a type II orientifold
compactification at the intersections of three stacks of D-branes with $U(3)\times U(2)\times U(1)$ gauge symmetry, that compactification's corresponding quiver
must be one of the mentioned 48, the couplings of which we can say a great deal about based on their global charges.

Popular avenues for studying quivers include systematic study of hypercharge embeddings and MSSM quivers \cite{Anastasopoulos:2006da}, as well as the study of particular quivers at the level of couplings \cite{Ibanez:2008my}. Additionally, systematic work has been done along these lines at the level of couplings \cite{Cvetic:2009yh} and the mass hierarchical structure of MSSM quarks and leptons have been investigated in \cite{Anastasopoulos:2009mr,Cvetic:2009ez}. The basic strategy in the systematic works was to study the phenomenology of MSSM quivers, possibly
extended by three right-handed neutrinos or a singlet $S$ which can give rise to a dynamical $\mu$-term. All presented quivers satisfy the necessary constraints
on the chiral spectrum for tadpole cancellation and a massless hypercharge, and thus it is not ruled out that these quivers can be embedded in a consistent
type II orientifold compactification. Beyond these necessary constraints, the quivers also satisfy a host of phenomenological constraints. In particular, 
for a quiver to be semi-realistic, one has to require that D-instanton effects generate enough of the forbidden Yukawa couplings to ensure that there are
no massless quark or lepton families. In doing so, however, a D-instanton which generates a Yukawa coupling might also generate a phenomenological
drawback, such as an R-parity violating coupling, a dimension five proton decay operator, or a $\mu$-term which is far too large. Such a quiver would be
ruled out.

\section{Conclusion and Outlook}
In these lectures, we have presented the basic perturbative and non-perturbative physics of type II orientifold compactifications. This corner
of the string landscape is particularly nice for understanding aspects of four-dimensional particle physics, as spacetime filling D-branes give
rise to four-dimensional gauge symmetry and chiral matter can appear at their intersections. The importance of the non-perturbative D-instanton
effects cannot be overstated in these compactifications, as in their absence a type II orientifold compactification often gives rise to
massless families of quarks or leptons, due to global $U(1)$ selection rules that forbid their Yukawa couplings. Indeed, they are \emph{necessary}
for some aspects of particle physics, as, for example, the Majorana neutrino mass term and the $10\,10\, 5_H$ top-quark Yukawa coupling are
\emph{always} perturbatively forbidden in weakly coupled type II. In addition to generating phenomenologically desirable couplings,
instanton effects must be taken into account because they could also generate couplings which spoil the physics. Furthermore, instantons
generate the leading superpotential contributions for K\" ahler moduli in type IIb, and thus play an important role in their stabilization.

Though the picture of particle physics in these models is beautiful, one still has to decide how to deal with the enormity of the landscape.
It is a useful fact that coupling issues can be studied
at the quiver level, which specifies how chiral matter transforms under the gauge groups of the D-branes, and thus the global $U(1)$ charges
of matter, while postponing the issue of global embeddings to a later date. We believe this approach to be of great use in identifying
promising quivers for global embeddings, as it is not worth trying to realize a global embedding for the sake of particle physics if it
can already be seen at the quiver level that a model is ruled out phenomenologically. In addition to phenomenological constraints on quivers,
string consistency conditions on chiral matter greatly constrain the possibilities, so it is not true that ``anything goes".

From the point of view of globally consistent type II orientifold compactifications, both type IIa and type IIb have their advantages
and disadvantages. On
one hand, the appearance of chiral matter at the intersection of D6-branes in type IIa is purely dependent on geometric issues, giving
an intuitive picture of particle physics. In addition, many useful CFT techniques have been developed for the type IIa string compactified
on a toroidal orbifold. In type IIb, on the other hand, the appearance of chiral matter depends on the choice of worldvolume flux
on the D7-branes, and therefore depends on more than geometry. The major advantage of type IIb, however, is that much more is known about moduli stabilization and that the
full power of complex algebraic geometry can be utilized, since the 1/2-BPS gauge branes and euclidean D-instantons wrap holomorphic divisors
rather than special Lagrangians. In addition, type IIb offers a description as the $g_s\mapsto 0$ limit of F-theory, which has been
of great interest in string phenomenology over the last few years\footnote{For work on F-theory GUTs, see \cite{Donagi:2008ca,Beasley:2008dc,Beasley:2008kw}. For lectures and further information, see \cite{Weigand:2010wm} and references therein.}.

The interplay between type IIb D-braneworlds and F-theory compactifications runs deep, as one might expect. In particular,
a major motivating factor for the study of the appearance of Georgi-Glashow GUTs in F-theory is the absence of the $10\, 10\, 5_H$ Yukawa
coupling in type IIb string perturbation theory. Though this coupling can be generated by an instanton in type II \cite{Blumenhagen:2007zk,Blumenhagen:2008zz}, it is exponentially
suppressed by the classical action of the instanton and therefore still has trouble explaining the top-quark hierarchy, unless the K\" ahler
moduli are stabilized such that the Yukawa coupling is $\cO(1)$.
In the F-theory lift of a type IIb GUT, the $10\, 10\, 5_H$ occurs at a point of $D_6$ enhancement, which is the lift of where the orientifold,
$U(5)$ brane, and $U(1)$ brane intersect in type IIb. These objects also can intersect at a point of $E_6$ enhancement in F-theory, which
gives a perturbative ($\cO(1)$) contribution to $10\, 10\, 5_H$. Though this is one of the phenomenological advantages of F-theory GUTs over type II, instantons still play important roles in other aspects of the physics. This is still an active area of research, and in particular the microscopic description of charged modes in F-theory needs clarification.

\acknowledgments

We would like to acknowledge I\~naki Garc\' ia-Etxebarria, Paul Langacker, Robert Richter and Timo Weigand for recent collaborations. We thank I.G.E. and R.R. in particular for discussions related to the content of the lectures. We thank the TASI organizers for providing a wonderful school, and the participants for lively discussions.  This work was supported in part by
the National Science Foundation under Grant No. NSF PHY05-51164, DOE under grant
DE-FG05-95ER40893-A020, NSF RTG grant DMS-0636606, the Fay R. and Eugene L.
Langberg Chair, and the Slovenian Research Agency (ARRS).

\clearpage



\bibliographystyle{JHEP}
\bibliography{refscheck}

\end{document}